\documentclass[lettersize,journal]{IEEEtran}

\usepackage[usenames, dvipsnames]{color}
\usepackage{color}
\usepackage{xcolor}
\usepackage{flafter}
\usepackage{placeins}
\usepackage{cite}
\usepackage{amsmath,amssymb,amsfonts}
\usepackage{algorithm}
\usepackage{algorithmic}
\usepackage{graphicx}
\usepackage{textcomp}
\usepackage{tabularx,array}
\usepackage{threeparttable}
\usepackage{mathrsfs}
\usepackage{soul}
\usepackage{setspace}
\usepackage{listings}
\usepackage{multicol}
\usepackage{multirow}
\usepackage{booktabs}
\usepackage{tikz}
\usepackage{rotating}
\usepackage{amsfonts}
\usepackage{hyperref}
\usepackage{ulem}

\usepackage{cite}
\usepackage{amsmath,amssymb,amsfonts}
\usepackage{algorithm}
\usepackage{algorithmic}
\usepackage{textcomp}
\usepackage{xcolor}
\usepackage{multicol}
\usepackage{multirow}
\usepackage{subcaption}
\usepackage{mathrsfs}
\usepackage{mathrsfs}
\usepackage{soul}
\usepackage{setspace}
\usepackage{colortbl}
\usepackage[dvipsnames]{xcolor} 
\usepackage{lastpage}
\usepackage{graphics}
\usepackage{wrapfig,lipsum,booktabs}
\usepackage{comment}
\usepackage{mathtools}

 \usepackage{array}
 \usepackage{threeparttable}
 \usepackage{xcolor}
 \makeatletter
 \newcommand*{\@rowstyle}{}
\newcommand*{\rowstyle}[1]{
 \gdef\@rowstyle{#1}%
 \@rowstyle\ignorespaces%
}
\newcolumntype{=}{
>{\gdef\@rowstyle{}}%
}
\newcolumntype{+}{
>{\@rowstyle}%
}
\newcolumntype{C}[1]{>{\centering\arraybackslash}p{#1}}
 \makeatother

\begin{document}

\title{Robust Multimodal mmWave Beam Prediction under Sensor Degradation}

\author{Saba M. Hosseini, Katarina Vuckovic, Amirhossein Safari, and Nazanin Rahnavard,~\IEEEmembership{Senior Member,~IEEE}

\thanks{S. M. Hosseini, A. Safari, and N. Rahnavard are with the ECE Dep., University of Central Florida, Orlando, FL 32816, USA (emails:  \{saba.mhosseini, amirhossein.safari, nazanin.rahnavard\}@ucf.edu).}
\thanks{This work was partly done when K. Vuckovic was with the ECE Dep., University of Central Florida, Orlando, FL 32816, USA (email: kvuckovic993@gmail.com).}%
}

\IEEEaftertitletext{\vspace{-.6cm}} 

\maketitle

\vspace{-3mm}
\begin{abstract}
Multimodal sensing-assisted deep learning has increasingly been adopted for mmWave beam prediction (BP), improving communication performance over conventional non-learning methods. However, practical fusion models remain vulnerable to over-reliance on dominant modalities and missing or degraded sensor inputs at inference. To address these challenges, we propose a robust low-complexity multimodal beam prediction framework that fuses vision, LiDAR, and GPS while maintaining performance under sensor failures. We train and evaluate late-fusion networks under clean and stressed sensing conditions, including missing, partially degraded, structured hybrid, and out-of-range degradations that are harsher than those seen during training. Modality-specific Auxiliary Supervision (AS) strengthens branch-level feature learning and reduces modality imbalance. Stochastic Robustness Training (RT) randomly removes or partially degrades sensor inputs, or their extracted features, during training to reduce reliance on individual sensors and improve cross-modal fusion. Experimental results show that the proposed model outperforms early-fusion, single-modality, and state-of-the-art benchmark approaches while maintaining low complexity and practical inference time, reducing beam-search overhead. It also achieves the best clean-condition performance and remains robust under sensor failures, whereas its counterpart without AS and RT degrades substantially. Finally, the best model reduces single-modality dominance by limiting the influence of any individual modality on the correct-beam prediction probability and reduces link outages caused by incorrect predictions over temporally continuous samples.

\vspace{-2mm}
\end{abstract}

\begin{IEEEkeywords}
mmWave Beam Prediction, Multi-modal Deep Learning, Sensor Failure, Robust Fusion, Auxiliary Supervision
\vspace{-2mm}
\end{IEEEkeywords}

\vspace{-2mm}
\section{Introduction}
Millimeter-wave (mmWave) communication is a key enabling technology for 5G and beyond systems, offering wide bandwidths at high carrier frequencies to support ultra-high data rates. Its short wavelengths allow dense antenna arrays and highly directional beamforming, but also make mmWave links sensitive to severe path loss and weak obstacle penetration. These challenges make accurate beam alignment essential for reliable connectivity~\cite{9791349}.

Accurate mmWave Beam Prediction (BP), which estimates the optimal transmitter--receiver beam direction, is critical for improving link performance and reducing energy consumption. However, BP remains challenging in dynamic environments, where user mobility, changing propagation conditions, and environmental variations can alter the optimal beam direction between the Base Station (BS) and the user over time. To resolve this challenge, multimodal sensing-aided BP is a promising solution. Integrating heterogeneous sensors into BP can improve situational awareness and guide beam selection effectively. In this paper, we propose a robust multimodal sensing-assisted Deep Learning (DL) BP framework for mmWave communication in dynamic environments.

We consider a realistic single-user scenario with predefined codebook beamforming without spatial multiplexing, where a vehicle equipped with Global Positioning System (GPS) communicates with a stationary BS equipped with vision, Light Detection and Ranging (LiDAR), and GPS sensors. To align with practical mmWave beam management, the proposed framework formulates BP as a discrete beam-index classification problem over a predefined beam codebook.
Using complimentary information from vision, LiDAR, and GPS, the framework aims to reduce beam-search overhead and support sensing-aided BP in dynamic environments without relying on slow, exhaustive RF-only probing at inference time.
 
Although multimodal fusion can improve prediction accuracy, fusion models may over-rely on dominant modalities and underutilize weaker sensor branches. Moreover, multimodal models are not inherently robust to missing or degraded sensors, creating a gap between controlled experimental settings and real-world deployment, where multimodal data quality is often uncertain and variable. Therefore, robustness to sensor degradation and unavailability is essential for practical multimodal BP.

To address these challenges, we employ \textit{late fusion}, in which each modality is processed by a dedicated encoder first. The extracted features are then concatenated and passed to a classification network. We investigate two complementary training strategies.
First, we introduce stochastic \textit{Robustness Training (RT)}, a family of training strategies that exposes the model to randomly selected clean and stressed sensor conditions. 
Under RT, we study two variants: (i) \textit{Robustness Training for missing modality (RT\textsubscript{M})}, which randomly uses either clean inputs, where all sensor modalities are available, or missing-modality cases generated by entirely suppressing one sensor modality. The proposed RT\textsubscript{M} improves robustness to complete modality unavailability in fusion models through missing-modality masking. (ii) \textit{Robustness Training for general degradation (RT\textsubscript{G})}, which generalizes RT\textsubscript{M} by randomly assigning each training sample to one of four sensor conditions: clean inputs, fully-missing modalities, partial sensor degradations, and hybrid cases that combine full missing and partial degradation.
Second, we employ \textit{Auxiliary Supervision (AS)}, which augments the late-fusion objective with modality-specific losses to encourage each branch to learn discriminative representations and improve fusion balance. Thus, while RT explicitly encourages the model to exploit the remaining reliable sensors when other modalities are degraded or unavailable, AS improves branch-level learning.

In our previous work~\cite{KatSabaMMBP1}, we presented preliminary results of our BP model and introduced comprehensive performance metrics. 
In this paper, we improve the robustness of fusion models using complementary AS- and RT-based strategies. Together, they improve fusion quality and resilience to sensor failures, provide interpretable modality-contribution behavior, and reduce outages across temporally consecutive samples. Robustness is learned through \textit{stress training} without dataset expansion, and is evaluated under missing, corrupted, hybrid missing-and-corrupted, and out-of-range degradations that are harsher than those seen during training. We also assess practical deployment aspects, including model complexity and inference time. To the best of our knowledge, this is the first study to analyze the robustness of AS- and RT-based multimodal BP under both missing- and corrupted-sensor conditions.
Our key contributions are:

\begin{itemize}
\item We propose an end-to-end multimodal late-fusion BP framework to effectively integrate heterogeneous modalities, outperforming unimodal and early-fusion baselines.

\item We introduce AS into models to strengthen modality-specific feature learning and improve fusion-stage decision reliability.

\item We develop a sensor-robust training framework under complete modality-level missingness, partial sensor degradation, and hybrid stress conditions.

\item Our simulations demonstrate improved resilience to sensor failures while providing insight into modality-contribution behavior. We validate our architecture against state-of-the-art methods, showing a favorable accuracy--complexity tradeoff.

\end{itemize}

The remainder of this paper is organized as follows. Section~\ref{Related Work} reviews related work. Section~\ref{System Model} presents the system model, problem formulation, performance metrics, experimental setup, and data preprocessing. Section~\ref{Network Structure} describes the proposed network architecture and RT. Section~\ref{Simulation} details training and evaluation. Section~\ref{Results} presents the results.

\vspace{-2mm}
\section{Related Work}
\label{Related Work}

\subsection{Beam Prediction (BP)}
Conventional mmWave beam selection relies on exhaustive or structured searches over large beamforming codebooks, requiring real-time pilot measurements across candidate beam pairs. This process introduces substantial overhead and latency, especially in mobile or latency-sensitive Multiple-Input Multiple-Output (MIMO) systems with large antenna arrays. At low pre-beamforming Signal-to-Noise Ratio (SNR), pilot measurements may also fail to reliably distinguish the optimal beam from the non-convex beam alignment problem. Proposed solutions, including compressive sensing methods address this issue via multiple-channel measurements, yet they remain sensitive to noise and environmental mismatches~\cite {10826962}.

DL-based methods address these limitations by shifting complexity of optimal BP from online search at communication time to offline training. DL-based BP can exploit complementary information, including environmental sensor data, to reduce or eliminate beam sweeping at inference, enabling scalable, low-latency operation in dynamic vehicular and urban scenarios. Early work used sub-6 GHz Channel State Information (CSI) for mmWave BP~\cite{alrabeiah2020deep,9743549}, while sensing-aided methods began with Global Positioning System (GPS) data~\cite{va2017position,morais2022position}, later expanding to vision~\cite{alrabeiah2020millimeter,Alkhateeb1,sarker2021vision,nishio2021wireless,ahn2022machine}, Light Detection and Ranging (LiDAR)~\cite{Klautau1,IPC,MBTVT,wu2022lidar}.

Recognizing the limitations of each single modality for BP, recent work has shifted to multimodal fusion to improve accuracy and robustness \cite{reviewNathan}. Late-fusion \cite{9796865,10279177,9764610,BeamBench,10683225,10577431,9771835,9751514,10464587}, hybrid-fusion \cite{tian2023multimodal}, and dynamic-fusion \cite{10912462} models fuse LiDAR, camera, and GPS for vehicular BP. A pilot-free four-modal late-fusion framework is presented in \cite{10683225}, but is limited to one night scenario, lacks cross-environment evaluation, and relies on a computationally intensive ResNet-50 image encoder. 
Multimodal BP using quantum–transformer networks \cite{10577431} achieved best performance via fusing GPS, LiDAR, and RGB features with a complex pipeline incorporating You Only Look Once (YOLO) and a vision transformer.
A Vision–language model-driven multimodal BP framework \cite{wang2025visionlanguagemodelvlmmeets} achieves marginal improvements in BP accuracy; however, its large model size and sensitivity to codebook size pose deployment challenges. 
\cite{tian2023multimodal} proposes a hybrid-fusion transformer that evaluates modality combinations and preprocessing strategies, reporting best results with camera+GPS fusion. These studies underline the accuracy-complexity trade-off in BP and the role of multimodal approaches for improving this trade-off.

\vspace{-2mm}
\subsection{Auxiliary Supervision (AS)}
Deep supervision, also called auxiliary classifier heads or Auxiliary Supervision (AS), is an effective strategy for training multimodal fusion models~\cite{LI2026132193}. By attaching supervised prediction heads to branch-level representations, each modality receives direct gradient signals instead of relying only on the fused output. This improves optimization and feature discriminability while reducing common multimodal issues such as overfitting, modality imbalance, and modality laziness, where dominant modalities suppress weaker branches.

In late fusion, AS is typically implemented by adding unimodal prediction heads and optimizing a weighted combination of the fused loss and the auxiliary unimodal losses. Recent studies show that AS preserves discriminative modality-specific representations during joint training and improves fusion quality~\cite{pmlr}, with applications in medical diagnosis, sentiment analysis, and emotion recognition~\cite{10230356,9156420,wu2024}.

\vspace{-2mm}
\subsection{Robust Multimodal Learning}
Sensor degradation broadly refers to reduced modality fidelity and availability, including partial sensor-data corruption and complete modality loss. In practice, multimodal models are not inherently robust to degraded inputs from sensor damage, temporary obstructions, noise, or environmental factors. Robust training strategies, such as sensor dropout \cite{mohta2020investigating}, address this by reducing overreliance on a single modality by exposing the model to sensor failures in training. 
Unlike the missing-modality BP approach in~\cite{11089951}, which expands the training set with masked or imputed samples, our RT framework samples degradation patterns in training and applies them directly to the original input samples, without increasing the dataset size.
\subsubsection{Robustness Training for Missing Modality (RT\textsubscript{M})}
Complete sensor failure represents a severe sensing disruption in which a modality becomes entirely unavailable; for example, camera damage may result in all camera frames missing. Sensor Dropout during training~\cite{pmlr,2670313} is a multimodal robustness strategy that randomly removes one or more sensor inputs during training, forcing the model to rely on the remaining modalities for prediction. 
This robustness adaptation is beneficial when sensor malfunctions, delays in sensor acquisition or data transmission, communication interruptions, power loss, or physical damage render a modality temporarily or permanently unavailable.
Prior work in multi-sensor prediction system has shown that sensor dropout improves robustness under missing-sensor conditions.
AMBER~\cite{wen2026} addresses robust multimodal BP under missing-modality conditions using a missing-modality-aware Transformer. Its strongest reported setting uses all available inputs, i.e., image, LiDAR, GPS, and historical beam-index information.

\subsubsection{Robustness Training for General Degradation (RT\textsubscript{G})}
Unlike complete missing-modality cases, in which we have no access to a given modality, in partial degradation, we still have access to corrupted portions of an modality. Partial degradation better reflects unpredictable real-world sensing impairments~\cite{liao2025benchmarking}, such as camera glare, adverse illumination, noise, blur, or partial occlusion, where measurements are degraded rather than absent. Prior multimodal robustness studies \cite{10588664,10209007} evaluate such failure modes separately, including noisy inputs, weather-induced corruptions, and calibration errors.

\vspace{-3mm}
\section{System Model and Problem Formulation}
\label{System Model}
\vspace{-1mm}

\subsection{Wireless Communication System Model}
\begin{figure}[!t]
    \centering
    \includegraphics[width=0.4\textwidth]{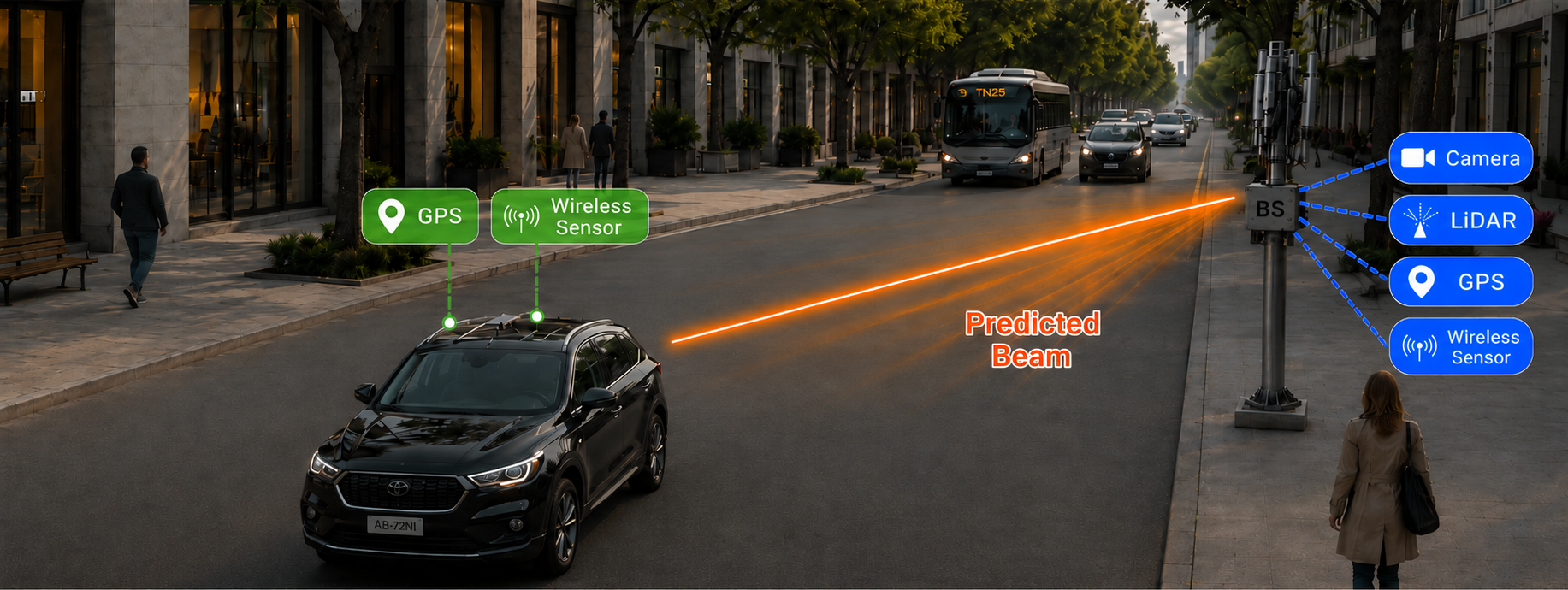} 
     \vspace{-1.5mm}
   \caption{\small{Illustration of the studied V2I multimodal system.}}
    \label{fig:StreetView}
    \vspace{-5mm}
\end{figure}

\vspace{-2mm}
\label{Communication Model}
We adopt the wireless system model in~\cite{charan2022multimodal} (Fig.~\ref{fig:StreetView}), where a BS equipped with a Uniform Linear Array (ULA) of $M$ mmWave antennas and multimodal sensors serves a single mobile user with an omnidirectional antenna and GPS. BS sensors include an RGB camera, LiDAR, and GPS. Communication follows a single-user orthogonal frequency-division multiplexing (OFDM) over $K$ subcarriers with codebook-based directional beamforming. Spatial multiplexing is not considered; therefore, although the BS is multi-antenna, the system differs from traditional MIMO settings that multiplex users or data streams.

Let $\mathbf{F}=\{\mathbf{f}_q\}_{q=1}^{Q}$ denote the beamforming codebook, where $\mathbf{f}_q\in\mathbb{C}^{M\times 1}$ is the $q$-th beamforming vector and $Q=64$ is the number of candidate beams. Let $\mathbf{h}_k[t]\in\mathbb{C}^{M\times 1}$ be the BS-side channel vector at subcarrier $k$ and time index $t$, where $k=1,\dots,K$. When the BS applies beam $\mathbf{f}_{q[t]}$, with $q[t]\in\{1,\dots,Q\}$ denoting the selected beam index at time $t$, the received signal at subcarrier $k$ is
\[
\scalebox{0.85}{$\displaystyle
y_k[t] = \mathbf{h}_k^{\mathsf{T}}[t]\mathbf{f}_{q[t]}x + v_k[t],
$}
\]
where $y_k[t]$ is the received signal, $x$ is the transmitted symbol satisfying $\mathbb{E}[|x|^2]=P$, $P$ is the transmit power, and $v_k[t]\sim\mathcal{CN}(0,\sigma^2)$ denotes circularly symmetric complex Gaussian receiver noise with variance $\sigma^2$. This received-signal model and the codebook-based beam selection follow~\cite{charan2022multimodal}.

\vspace{-2mm}
\subsection{Beam Prediction Formulation and Evaluation Metrics}
\label{Beam Predicition Formulation}
We formulate beam selection as a codebook-based prediction task, where the model selects the best beam index from a predefined codebook rather than designing beamforming vectors. This reflects practical mmWave systems, in which feasible beams are constrained by hardware and implementation requirements, making codebook-based DL more deployable than codebook-free beamforming.

\subsubsection{Formulation} 
The goal of BP is to select, at time \(t\), the beam index \(q[t]\) that maximizes the average effective channel gain $G_q[t]$ across the  \(K\) subcarriers given by:
\vspace{-1.7mm}
\[
\scalebox{0.85}{$\displaystyle
G_q[t] \triangleq \frac{1}{K}\sum_{k=1}^{K}\left|\mathbf{h}_k^{\mathsf{T}}[t]\mathbf{f}_q\right|^2,
$}
\]

The optimal beam label is therefore given by
\[
\scalebox{0.85}{$\displaystyle
q^{*}[t] = \operatorname*{arg\,max}_{q \in \{1,\ldots,Q\}} G_q[t].
$}
\]

The optimal beam index is typically obtained using explicit channel information, which is difficult to acquire in mmWave systems, or through exhaustive search.

Rather than relying on a single sensing snapshot, the proposed framework predicts the optimal beam at time $t$ using five consecutive camera and LiDAR observations at time steps $t-4,\ldots,t$, together with two GPS readings available at the first two time steps, $t-4$ and $t-3$. Let $\mathbf{v}[\tau]$, $\mathbf{l}[\tau]$, and $\mathbf{g}[\tau]$ denote the vision, LiDAR, and GPS observations at time step $\tau$, respectively. The predicted beam index is formulated as
\[
\scalebox{0.85}{$\displaystyle
\hat{q}[t] = F_{\boldsymbol{\theta}}\left(
\{\mathbf{v}[\tau],\mathbf{l}[\tau]\}_{\tau=t-4}^{t},
\{\mathbf{g}[t-4],\mathbf{g}[t-3]\}
\right),
$}
\]
where $F_{\boldsymbol{\theta}}$ denotes the proposed learning model. The model processes features extracted from consecutive observations using a GRU to capture short-term temporal dependencies and predict a single beam index at time $t$.

\subsubsection{Metrics} 
\label{Beam Prediction Eval Metrics} 
In mmWave BP, model performance cannot be adequately characterized by a single metric. Therefore, we adopt multiple metrics from our prior work \cite{KatSabaMMBP1} to provide a comprehensive and meaningful evaluation of the proposed method.
Top-$K$ Accuracy (top-$K$A) measures whether the true beam index appears among the $K$ highest-probability predictions.
Distance-Based Accuracy (DBA) Score \cite{charan2022multimodal} quantifies the proximity between true and predicted beams, with higher scores indicating closer predictions.
Weighted multi-class precision (P) evaluates classification performance under class imbalance by averaging the class-wise precision values, with each class weighted according to its relative frequency in the dataset~\cite{alkhateeb2023deepsense}.
Top-K Beams (top-KB) evaluates BP by checking if the predicted beam fall within the highest-power ground-truth beams.  
Power Ratio (PR) quantifies proximity of predicted beam’s power to the optimal beam’s power. We also report inference time of the model per sample.

\vspace{-2mm}
\subsection{Experimental Scenarios Setup}
The datasets used here are drawn from DeepSense6G Scenarios~32--34~\cite{charan2022multimodal,alkhateeb2023deepsense}, which capture real-world outdoor 60\,GHz Vehicle-to-Infrastructure (V2I) mmWave deployments in dynamic urban environments during day and night. BS is fixed and equipped with a 16-element ULA and a 64-beam codebook. It is co-located with an RGB camera and a 3D LiDAR, while the mobile user carries a quasi-omnidirectional 60\,GHz transmitter and an onboard GPS receiver. As the user moves through street-level scenes containing buildings, vehicles, pedestrians, and other static or dynamic obstructions, the BS records time-aligned sensing data and the user's transmitted signal. These conditions affect beam selection through vehicle motion, surrounding geometry, and object blockage, reflecting practical BP settings with multiple objects, partial positional information, and possible delayed or missing user-side information.

Each sample contains \textit{five consecutive} BS-side sensing frames from vision and LiDAR, \textit{two} GPS readings of the user's location, and the corresponding top-1 optimal beam index from the 64-beam codebook. Beam power distributions across the 64 codebook indices are also provided. As noted in~\cite{alkhateeb2023deepsense}, these dimensionless values represent relative, rather than absolute, power. They are not normalized to a fixed maximum, but reliably capture relative differences, enabling meaningful intra-dataset comparisons without direct interpretation in absolute units such as Watts or dBm. Since BP is meaningful only when a viable link exists, samples labeled as fully blocked by the beam-power criterion in~\cite{hosseini2027blockage} are excluded, as full-blockage is handled by a separate detector in that work. Accordingly, the proposed algorithms are sensing-aided, single-user, codebook-based BP methods for urban outdoor deployments, using BS-side vision and LiDAR sensing together with limited user-side GPS information.

\vspace{-3mm}
\subsection{Data Preprocessing}
\label{Data  Preprocessing}
Here, we describe preprocessing pipelines for modalities. 
\subsubsection{LiDAR (L)} 
Raw LiDAR point clouds are converted into Bird’s Eye View (BEV) images by filtering and projecting 3D points onto a 2D grid. Each point, with \((x,y,z)\) coordinates and intensity, undergoes two filtering steps:  
1) \textit{Azimuth filter:} Retains points with azimuth angle  \( \in [110^\circ, 360^\circ]\), restricting the field of view to the region of interest.  
2) \textit{Spatial filter:} Limits $x$ and $y$ to $[-60, 40]$~m, selecting
an area of $100m\times100m$.  The filtered points are projected onto a \(150\times150\)-pixel grid to produce a BEV RGB image as 
red, green, and blue encoding height ($z$) (vertical structure), intensity (material properties), and density, i.e., the normalized number of points per grid cell ([0, 1]), highlighting object surface details, correspondingly. This representation combines height, intensity, and density to provide a comprehensive view of the scene’s structure. 

\subsubsection{Camera Vision (V)}
RGB camera frames of size \(540 \times 960 \times 3\) are spatially downsampled to \(150 \times 150 \times 3\) to reduce redundancy, ensure dimensional consistency with the other modalities, and lower computational cost.

\subsubsection{GPS (G)}
GPS latitude–longitude measurements are converted to Cartesian coordinates and normalized relative to the BS position, as in \cite{KatSabaMMBP1, vuckovic2024robust}. Because only the first two readings are available in each sequence, three additional samples are extrapolated to create five-step sequences consistent with other modalities. GPS calibration, as in \cite{tian2023multimodal, BeamBench}, aligns coordinate systems across scenarios by standardizing the angular orientation of beam 32, ensuring spatial and angular consistency.  

For early fusion, user positions are encoded as \((150,150)\) binary maps, where the cell corresponding to
the point in Cartesian coordinates is set to 1 and all others to 0, ensuring consistent dimensions across modalities. This map-based encoding is specific to early fusion and is not applied in the GPS-only or late-fusion models.

\vspace{-2mm}
\subsection{Data Normalization}
Normalization is applied to mitigate scale disparities across modalities. For the single-vision model, pixel values are scaled to $[0,1]$ by dividing by 255. In early-fusion models, vision (3 channels) and LiDAR (3 channels) features are concatenated and L2-normalized across channel dimension so each spatial–temporal vector has unit Euclidean norm. GPS data, represented as a highly sparse 2D map, is excluded to avoid skewed results and concatenated subsequent to other modalities normalization. This balances modality contributions and prevents dominance by features with larger ranges.

\vspace{-2mm}
\section{Proposed Beam Prediction Framework}
\label{Network Structure}
Our BP framework uses preprocessed multimodal sensor data to predict the optimal beam without requiring Radio Frequency (RF) measurements or directional signal strength information during \textit{inference}. Instead of explicitly modeling the channel, it \textit{learns} implicit mappings between spatial--geometric environmental features and beam-selection outcomes. This provides a practical, channel-agnostic alternative for real-time mmWave operation, particularly when RF probing is costly or impractical.

\vspace{-2mm}
\subsection{Network Structure}
\begin{figure}[!t]
    \centering
    \begin{subfigure}[b]{0.4\textwidth}
        \centering
        \includegraphics[width=0.75\textwidth]{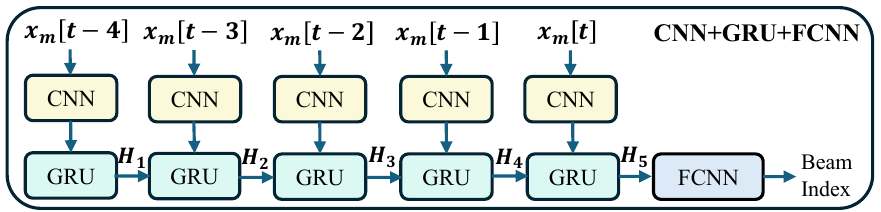} 
        \vspace{-2mm}
        \caption{}
        \label{fig:CNN_GRU_FCNN}
    \end{subfigure}
    \hfill
    \begin{subfigure}[b]{0.4\textwidth}
        \centering
        \includegraphics[width=1\textwidth]{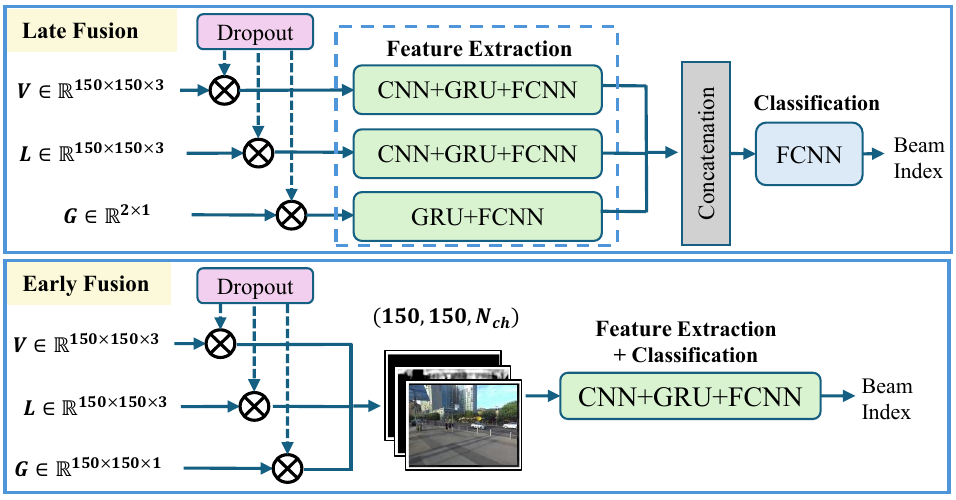} 
        \vspace{-5mm}
        \caption{}
        \label{fig:Early_and_Late_Fusion}
    \end{subfigure}
    \vspace{-2mm}
    \caption{\small{(a) Proposed CNN+GRU+FCNN architecture for the BP model, where \(x_{m}[t]\) is the modality $m$ input at time \(t\). (b) The overall input-output process in the late- and early-fusion configurations based on the architecture in (a), with the proposed modality dropout.}}
    \label{fig:overall_CNN_GRU_FCNN}
    \vspace{-4mm}
\end{figure}

The proposed BP network architecture comprises three modules: a Convolutional Neural Network (CNN) for spatial feature extraction, a Gated Recurrent Unit (GRU) for temporal modeling, and a Fully-Connected Neural Network (FCNN) for feature integration and classification. These modules can operate on single-modality inputs or be extended to fusion architectures for multimodal data. BP models adopt a CNN+GRU+FCNN architecture (Fig.~\ref{fig:CNN_GRU_FCNN}), suitable for dynamic environments and allowing robust, high-accuracy predictions. The structure of each module is described below. 

\textbf{CNN:}
The CNN module extracts frame-level spatial features from vision and LiDAR. It uses a stack of 2D convolutional layers with kernel sizes decreasing from {\small$7\times7$} to {\small$3\times3$} and filter counts increasing from 4 to 256. Each convolutional layer is followed by ReLU activation, batch normalization with momentum 0.9, and max pooling.
BP models employ six (five in late fusion) convolutional layers for vision and LiDAR, while GPS inputs, containing only two scalar values, bypass CNN processing in single-modality GPS and the GPS branch of late-fusion models. 

\textbf{GRU:}  
The GRU module captures temporal correlations of spatial features across frame sequences. 
It consists of a single 64-unit layer with 0.8 dropout.

\textbf{FCNN:}  
The FCNN integrates features from the GRU module, reduces dimensionality, and outputs class logits. Models use a single 64-unit dense layer with ReLU activation and 50\% dropout, except for the GPS model, which employs two dense layers with 128 and 64 units. The output is generated through a softmax layer over 64 beam classes.

\vspace{-4mm}
\subsection{Late Fusion}
\vspace{-1mm}
In late fusion (Fig.~\ref{fig:Early_and_Late_Fusion}, top), each modality is processed by a dedicated CNN+GRU+FCNN branch, and the extracted features are concatenated and passed to an FCNN for classification. This design supports varying modality dimensions and branch-specific tuning (e.g., bypassing CNN processing for GPS, which has only two values).

\textbf{Auxiliary Supervision:}
We augment the standard late-fusion objective with fixed-weight auxiliary-branch losses to stabilize optimization, improve generalization, mitigate modality imbalance, and accelerate convergence. The resulting objective jointly optimizes sensor-specific and fused predictions, encouraging each branch to learn discriminative beam-prediction features, while the fused head learns complementary multimodal representations.

Let \(\mathcal{M}=\{1,\ldots,N_m\}\) denote the set of available modalities, where \(N_m\) is the number of modalities and \(m\in\mathcal{M}\) indexes each modality. Each modality branch is supervised through both the final fused prediction and an auxiliary classification head. The feature extractor \(f_m\) maps \(\mathbf{x}_m\), the input of modality \(m\), to a latent representation $\mathbf{h}_m = f_m(\mathbf{x}_m)$.
The auxiliary head of each branch produces modality-specific logits and probabilities as
\begin{equation*}
\scalebox{0.85}{$\displaystyle
\mathbf{z}_m = \mathbf{W}_m \mathbf{h}_m + \mathbf{b}_m, 
    \qquad
    \hat{\mathbf{y}}_m = \mathrm{softmax}(\mathbf{z}_m),
$}
\end{equation*}
where \(\mathbf{W}_m\) and \(\mathbf{b}_m\) are the trainable weight matrix and bias vector of the auxiliary head for modality \(m\), \(\mathbf{z}_m\) denotes the logits, and \(\hat{\mathbf{y}}_m\) is the predicted class-probability vector. For the fusion output, the modality features are first normalized, concatenated, and passed through the fusion module \(f_{\mathrm{fus}}\):
\begin{equation*}
\scalebox{0.85}{$\displaystyle
    \mathbf{h}_{\mathrm{fus}} =
    f_{\mathrm{fus}}\left(
    \operatorname{Concat}_{m\in\mathcal{M}}
    \big(\mathrm{BN}(\mathbf{h}_m)\big)
    \right).
$}
\end{equation*}
where \(\mathrm{BN}(\cdot)\) denotes batch normalization, \(\operatorname{Concat}_{m\in\mathcal{M}}(\cdot)\) denotes concatenation over all modality branches, and \(\mathbf{h}_{\mathrm{fus}}\) is the fused latent representation.
The fused logits and prediction probabilities are obtained by
\begin{equation*}
\scalebox{0.85}{$\displaystyle
    \mathbf{z}_{\mathrm{fus}} =
    \mathbf{W}_{\mathrm{fus}}\mathbf{h}_{\mathrm{fus}} + \mathbf{b}_{\mathrm{fus}},
    \qquad
    \hat{\mathbf{y}}_{\mathrm{fus}} =
    \mathrm{softmax}(\mathbf{z}_{\mathrm{fus}}).
$}
\end{equation*}
where \(\mathbf{W}_{\mathrm{fus}}\) and \(\mathbf{b}_{\mathrm{fus}}\) are the trainable weight matrix and bias vector of the fusion classifier, \(\mathbf{z}_{\mathrm{fus}}\) denotes the fused logits, and \(\hat{\mathbf{y}}_{\mathrm{fus}}\) is the fused class-probability vector. Given the one-hot ground-truth beam label \(\mathbf{y}\), the auxiliary and fusion losses are defined using categorical cross-entropy:
\begin{equation*}
\scalebox{0.85}{$\displaystyle
    \mathcal{L}_m =
    - \sum_{c=1}^{N_c} y_c \log(\hat{y}_{m,c}),
    \qquad
    \mathcal{L}_{\mathrm{fus}} =
    - \sum_{c=1}^{N_c} y_c \log(\hat{y}_{\mathrm{fus},c}),
$}
\end{equation*}
where \(N_c\) is the number of beam classes, \(c\) indexes the beam class, \(y_c\) is the \(c\)-th element of \(\mathbf{y}\), and \(\hat{y}_{m,c}\) and \(\hat{y}_{\mathrm{fus},c}\) are the \(c\)-th predicted probabilities from the modality-specific and fused outputs, respectively. Despite the classification being conducted using only the output from the fusion network, the total training objective is
\begin{equation*}
\scalebox{0.85}{$\displaystyle
    \mathcal{L}_{\mathrm{total}} =
    \lambda_{\mathrm{fus}}\mathcal{L}_{\mathrm{fus}}
    +
    \sum_{m \in \mathcal{M}} \lambda_m \mathcal{L}_m,
$}
\end{equation*}
where \(\lambda_{\mathrm{fus}}\) and \(\lambda_m\) control the contributions of the fused and modality-specific supervision terms, respectively. 

\vspace{-2mm}
\subsection{Early Fusion} 
For comparison, we also study early fusion which concatenates all preprocessed modalities along the channel axis at the input stage, before feature extraction (Fig.~\ref{fig:Early_and_Late_Fusion}, bottom). 
Each sample is a five-step sequence of \(150 \times 150\) frames with \(N_{\mathrm{ch}}\) total input channels after channel-wise modality concatenation. 
This design captures cross-modal correlations from the outset, but enforces uniform input dimensions across modalities, which can limit representation and cause information loss \cite{Shankar}. 

\vspace{-2mm}
\subsection{Robustness Training}
Unavailable or degraded sensor readings can directly reduce the reliability of link predictions. For robustness to such failures, we implement two \textit{training strategies}: RT\textsubscript{M} and RT\textsubscript{G}. At inference, inputs are passed unchanged unless an explicit degraded-condition evaluation is performed.

\subsubsection{RT\textsubscript{M}}
During training, each sample is assigned either a clean-input case with no modality dropout or degradation, with probability \(1-p_f\), or a single-modality dropout case with probability \(p_f\). In the dropout case, one of the \(N_m\) available modalities is selected uniformly, giving each modality a dropout probability of \(p_f/N_m\). Unlike pixel-, channel-, or feature-level suppression, RT\textsubscript{M} removes an entire input modality: for sample \(i\), a binary modality mask \({z}_{i}=[z_{i,1},\ldots,z_{i,N_m}]\) is generated, where \(z_{i,m}\in\{0,1\}\) indicates whether modality \(m\) is kept or dropped. If \(z_{i,m}=0\), the full input tensor of modality \(m\) is zeroed across all time steps and feature dimensions; otherwise, it is passed unchanged. This trains the fusion network on both complete and incomplete multimodal inputs, reduces over-reliance on any single sensor, and encourages the use of remaining reliable modalities without reconstructing, generating, or retrieving missing inputs.

In early fusion, RT\textsubscript{M} zeros the selected modality at the channel-group level before the fused raw input is passed to the shared feature extractor. In late fusion, RT\textsubscript{M} zeros the selected modality at the input of its corresponding modality-specific encoder. The affected branch therefore receives a zero input, while the embeddings from all modality branches are still concatenated and passed to the fusion classifier.

\subsubsection{RT\textsubscript{G}}
We adopt a more general RT variant, RT\textsubscript{G}, which exposes the fusion model to a mixture of clean samples, partially degraded inputs, complete missing-modality cases, and hybrid sensor conditions.  
For each training sample, the robustness layer selects one global sensor condition---clean, partial, missing, and hybrid---with probabilities \(p_c\), \(p_p\), \(p_m\), and \(p_h\), respectively, where {\small\(p_c+p_p+p_m+p_h=1\)}.
Accordingly, in expectation, a fraction \(p_c\), \(p_p\), \(p_m\), \(p_h\) of inputs remains unchanged, receives one partial modality degradation, has one modality fully removed, and one modality is partially degraded while another modality is fully removed, respectively.

Since Late LVG (the late-fusion model combining L, V, and G) achieves the best performance, we apply RT\textsubscript{G} only to this model.
For each degradation case, the affected modalities are sampled uniformly. In full missing-modality training, one modality {\small\(m_f \sim \mathrm{Uniform}\{\mathrm{V},\mathrm{L},\mathrm{G}\}\)} is selected and replaced with zeros within pixels of all channels and frames of that sample. In partial degradation training, one modality {\small\(m_p \sim \mathrm{Uniform}\{\mathrm{V},\mathrm{L},\mathrm{G}\}\)} is selected, and one corruption type is then sampled uniformly from the enabled degradation set of that modality. In the hybrid case, distinct modalities $m_f$ and $m_p$ are selected and the former is fully removed, while the latter is partially degraded.
The modality-specific partial degradation types considered in this work are summarized below.

For the image-like modalities, namely vision and LiDAR BEV, let
$v,v',l,l' \in [0,255]^{T\times H\times W\times C},$
where \(v\) and \(l\) denote the original vision and LiDAR sequences, and
\(v'\) and \(l'\) denote their corrupted counterparts. Here, \(T=5\) is the
number of time steps, \(H\) and \(W\) are the spatial dimensions, and \(C\) is
the number of channels. 

The indices {\small\(t\in\{1,\ldots,T\}\)}, {\small\(h\in\{1,\ldots,H\}\)}, {\small\(w\in\{1,\ldots,W\}\)}, and {\small\(c\in\{1,\ldots,C\}\)} denote the frame, row, column, and channel, respectively. For compactness, we write {\small\(i=(t,h,w,c)\)}, so that {\small\(v_i\)}, {\small\(v'_i\)}, {\small\(l_i\)}, and {\small\(l'_i\)} refer to the corresponding entries of {\small\(v\)}, {\small\(v'\)}, {\small\(l\)}, and {\small\(l'\)}. For intensity-modifying corruptions, such as noise, brightness increase, and fog, the corrupted values are clipped to the valid range using
{\small$\mathrm{clip}_{255}(z)=\min\{\max\{z,0\},255\}.$}

\textbf{Vision degradations.}
The partial degradations are \cite{10588664}:

\textit{Gaussian noise:}
Let \(\epsilon\) denote additive Gaussian noise with the same 
dimensions as \(v\). Noise is applied as \(v'=\mathrm{clip}_{255}(v+\epsilon)\), 
where each element of \(\epsilon\) is independently sampled from \(\mathcal{N}(0,\sigma_v^2)\). This simulates random image sensor noise.

\textit{Pixel dropout:}
At each pixel index {\small\((t,h,w)\)}, all relevant channels are zeroed with probability \(\rho^{\mathrm{pixel}}\).

\textit{Patch masking:}
Let {\small\(M^{\mathrm{patch}}\in\{0,1\}^{T\times H\times W}\)} denote a mask that is zero inside a random rectangular region and one elsewhere. Patch masking is applied as {\small$v' = v \odot M^{\mathrm{patch}}$}, with $\odot$ denoting element-wise multiplication. The patch height and width are {\small\(H_p=\lfloor \alpha H \rfloor\)} and {\small\(W_p=\lfloor \alpha W \rfloor\)}, respectively. The same patch location is applied across all five frames, simulating a temporally consistent occlusion from partial camera blockage.

\textit{Blur:}
Let {\small\(B\)} denote a {\small\(b \times b\)} averaging filter, where
{\small\(B=\frac{1}{b^2}\mathbf{1}_{b\times b}\)}. Blur is implemented as {\small\(v'=\mathrm{clip}_{255}(v * B)\)}, where \(*\) denotes 2D convolution applied independently to each frame and color channel. This simulates defocus or motion blur.

\textit{Brightness:}
The RGB image is converted to the hue--saturation--value (HSV) space, where the value channel represents pixel brightness. Brightness is increased by adding an offset {\small\(\Delta_b/255\)} to the value channel and clipping the result to {\([0,1]\)}. The  HSV image is converted back to RGB to obtain {\small\(v'\)}. This corruption simulates overexposure or strong illumination.

\textit{Fog:}
Let \(\lambda\textsuperscript{fog}\in[0,1]\) denote fog strength. Fog is simulated by blending a pixel with a white image as {\small\(v'_{i}=\mathrm{clip}_{255}\!\left((1-\lambda\textsuperscript{fog})v_{i}+\lambda\textsuperscript{fog}\cdot 255\right)\)}. This shifts pixel intensities toward white: dark pixels brighten, bright pixels remain near saturation, and intensity differences shrink. Thus, image contrast is reduced, approximating foggy or hazy conditions.

\textit{Temporal frame dropout:}
A random subset of frames is removed from the affected modality by setting them to zero, while the remaining frames remain unchanged. Dropped frames need not be consecutive. Per sample, the number of dropped frames is drawn from {\small\(\{1,2,3,4\}\)} with probabilities \(p_{t1}\), \(p_{t2}\), \(p_{t3}\), and \(p_{t4}\). This simulates missing frames in sequential sensor inputs and encourages the model to rely on available observations rather than assuming a complete sequence. Dropping all five frames is equivalent to complete modality removal and is handled as the missing-modality case.

\textbf{LiDAR degradations.}
The partial degradations are Gaussian noise, patch masking, temporal frame dropout, and sparsification. Gaussian noise and temporal frame dropout use the same definitions and parameters as in the vision-degradation case. Patch masking is applied similarly, and the masked region corresponds to a local LiDAR occlusion or spatial blind spot. Sparsification is LiDAR-specific and is described below.

\textit{Sparsification:}
Sparsification removes each occupied LiDAR cell with probability \(\rho^{\mathrm{sparse}}\). 
A cell at location {\small\((t,h,w)\)} is occupied if {\small\(\max_c |l_{t,h,w,c}|>10^{-6}\)}. 
Dropped cells have all channels zeroed; otherwise, they remain unchanged. This removes, on average, a fraction \(\rho^{\mathrm{sparse}}\) of occupied cells, simulating reduced LiDAR returns or a sparser BEV representation.

\textbf{GPS degradations.}
For GPS, the five-step sequence extrapolated from the original two frames is {\small\(g=[g[0],\ldots,g[4]]\)}, where {\small\(g[t]\in\mathbb{R}^2\)} is the 2D coordinate at time \(t\). Let \(g'[t]\) denote the degraded coordinate, and let {\small\(\mathcal{T}=\{0,\ldots,4\}\)} denote the temporal index set. The following degradations are considered.

\textit{Gaussian GPS noise:}
We model GPS noise as a sequence-level Gaussian localization bias shared across all time steps, {\small\(g'[t]=g[t]+b_n\)} for {\small\(t\in\mathcal{T}\)}, where {\small\(b_n\in\mathbb{R}^2\)}, {\small\(b_n\sim\mathcal{N}(0,{\epsilon_n}^2 I)\)}. The bias \(b_n\) models a consistent short-term GPS offset rather than independent frame-wise noise.

\textit{Drift:}
GPS drift is a time-dependent offset that increases along the sequence. A drift vector {\small\(d\in\mathbb{R}^2\)} is sampled per sequence as {\small\(d\sim\mathcal{N}(0,{\epsilon_d}^2 I)\)}, and applied as {\small\(g'[t]=g[t]+td\)} for {\small\(t\in\mathcal{T}\)}. Thus, the first point remains unchanged, while later points receive progressively larger offsets, approximating accumulated localization error over the extrapolated trajectory.

\textit{Frozen GPS:}
Frozen GPS repeats the first coordinate across the sequence, {\small\(g'[t]=g[0]\)} for {\small\(t\in\mathcal{T}\)}, simulating a position estimate that stops updating.

\vspace{-2mm}
\section{Simulation}
\label{Simulation}

\subsection{Network Training}
\label{Training}
Models are trained end-to-end and evaluated using stratified 10-fold cross-validation to address class imbalance and reduce variance in performance estimates. 
Each fold preserves class proportions, uses one fold for testing and the remaining nine folds for training, with 20\% of the training set reserved for validation. This protocol ensures that each sample is tested once and provides an unbiased, and low-variance performance estimate. 
Models were trained using Adam optimizer with learning rates of {0.0001, 0.001, 0.01}, a batch size of 30, and up to 300 epochs. Early stopping was used to prevent overfitting, and the learning rate was reduced by a factor of 0.1 after four epochs without validation accuracy improvement. \footnote{Code to be released: \url{https://github.com/SabaMh/MMBP2}}

Models classify the correct beam among 64 codebook entries. A custom monitoring process tracks both accuracy and PR, saving the model with the best PR to balance accuracy and signal strength, ensuring effective prediction. PR is computed on the top-\(k\) predicted beams after each epoch and NaN values are replaced with 0 to ensure robust PR calculations.

For RT, we first train baseline early- and late-fusion models with different sensor combinations under clean conditions. The best-performing RT\textsubscript{M} configuration (Late LVG) is then used for RT\textsubscript{G}. As an extended variant of RT\textsubscript{M}, we also train and evaluate a more aggressive RT\textsubscript{M(1/2)} variant of Late LVG. In this case, a single- or dual-modality dropout pattern is applied with probability \(p_f\). When dropout is applied, one valid pattern is selected uniformly from {\small$\{V,L,G,VL,VG, LG\}$}. 
Hence, each dropout pattern occurs with probability \(p_f/6\).
This exposes the model to both moderate and severe missing-sensor conditions during training. The AS and RT hyperparameters of the degradation-based training schemes are listed in Table~\ref{tab:robust_settings} and were selected experimentally to balance performance under complete- and degraded-data conditions. 

\newcommand{\smath}[1]{{\fontsize{6}{6.5}\selectfont $#1$}}
\newcommand{\smodel}[1]{{\fontsize{6.2}{6.5}\selectfont #1}}

\begin{table}[t]
\centering
\caption{\small{RT and AS hyperparameters.}}
\vspace{-2mm}
\label{tab:robust_settings}
\scriptsize
\setlength{\tabcolsep}{0.9pt}
\renewcommand{\arraystretch}{.9}

\begin{tabularx}{\columnwidth}{|>{\raggedright\arraybackslash}p{0.35\columnwidth}|>{\raggedright\arraybackslash}X|}
\hline
\textbf{Model} & \textbf{Hyperparameters} \\
\hline

\multicolumn{2}{|l|}{\textbf{RT\textsubscript{M} models}} \\
\hline

\smodel{Early LV (RT\textsubscript{M})}
& \smath{p_f=0.15}\\
\hline

\smodel{Late LV (AS+RT\textsubscript{M})}
& \smath{p_f=0.30;\ \lambda_V=1.3,\ \lambda_L=1.5,\ \lambda_{fus}=2}. \\
\hline

\smodel{Late VG (AS+RT\textsubscript{M})}
& \smath{p_f=0.20;\ \lambda_V=\lambda_G=0.8,\ \lambda_{fus}=1}. \\
\hline

\smodel{Late LVG (AS+RT\textsubscript{M})}
& \smath{p_f=0.15;\ \lambda_V=\lambda_L=\lambda_G=0.8,\ \lambda_{fus}=1}. \\
\hline

\smodel{Late LVG (AS+RT\textsubscript{M(1/2)})}
& \smath{p_f=0.15;\ \lambda_V=\lambda_L=\lambda_G=0.8,\ \lambda_{fus}=1}. \\
\hline

\multicolumn{2}{|l|}{\textbf{RT\textsubscript{G} model}} \\
\hline

\multicolumn{2}{|>{\raggedright\arraybackslash}p{\dimexpr\columnwidth-2\tabcolsep-2\arrayrulewidth\relax}|}{%
\smodel{Late LVG (AS+RT\textsubscript{G})}
\smath{(p_c,p_p,p_m,p_h)=(0.5,0.2,0.2,0.1);
(p_{t1},p_{t2},p_{t3},p_{t4})=(0.4,0.3,0.2,0.1);
\sigma=10,\ \rho^{\rm pixel}=0.10,\ \alpha=0.20,\ b=7,\ \Delta_b=50,\ \lambda\textsuperscript{fog}=0.35;
\rho^{\rm sparse}=0.40;\ \epsilon_n=0.2,\ \epsilon_d=0.04;
\lambda_V=\lambda_L=\lambda_G=0.8,\ \lambda_{fus}=1}.} \\
\hline
\end{tabularx}
\vspace{-3mm}
\end{table}

\vspace{-2mm}
\subsection{Evaluation Approach} 
All results are reported as the mean \(\pm\) standard deviation across folds. For late-fusion models trained with AS, evaluation uses only the final fusion output. We first assess the effect of AS and RT by comparing the performance of all fusion models with and without those. For RT evaluation, each failure condition is applied to all test samples to measure robustness under persistent sensor failure. Each robustness-trained model is tested using the same degradation type used during training.

We further analyze late-fusion models’ reliance on each modality to predict the true beam before and after adding AS and RT. For each modality, we measure the decrease in the predicted true-beam probability after ablating that modality:
\begin{equation}
\scalebox{0.85}{$\displaystyle
\Delta p_{\mathrm{true}}^{(m)}
=
p_{\mathrm{true}}(\text{all modalities})
-
p_{\mathrm{true}}(\text{modality } m \text{ ablated}),
\label{DeltaPTrue}
$}
\end{equation}
where \(m\) denotes the ablated modality and \(p_{\mathrm{true}}\) is the predicted probability of the true beam class. A larger \(\Delta p_{\mathrm{true}}^{(m)}\) indicates stronger reliance on modality \(m\) for supporting the correct beam prediction, whereas a smaller value indicates that removing modality \(m\) has a weaker effect.

To study partial test-time modality loss, we evaluate Late VG model, as a representative fusion model, when one modality is dropped for only a fraction of the test samples rather than for the entire test set.
We evaluate RT\textsubscript{G} across multiple individual degradation types, including conditions that are harsher than those seen during training. The harsher setting includes blurred V with \(b=40\), sparsified L with \(\rho^{\rm sparse}=0.7\), and drifted G with \(\epsilon_d=0.3\). To assess link reliability, we further evaluate RT\textsubscript{G} on a temporally continuous set of clean samples by measuring the longest run of consecutive successful predictions, defined with respect to top-\(k\)A.

\vspace{-2mm}
\section{Results and Discussion}
\label{Results}
This section evaluates the proposed models under both clean and sensor-impaired conditions. Clean-test results refer to evaluation using the original test inputs without modality dropout or degradation and are used to assess the baseline predictive performance of each model. We then analyze the effects of AS and the performance of RT variants under missing-modality and partial-degradation conditions.

\vspace{-2mm}
\subsection{Clean-Test Performance Analysis}
Table~\ref{ResultsTable} reports the complexity and clean-test performance of all models built on the proposed architecture. The metrics are reported as scenario-weighted averages, where each scenario is weighted by its number of test samples. 

\begin{table}[t]
\centering
\caption{\small{Complexity and performance of the proposed classifiers with and without AS and 
RT on test samples. Metrics: trainable parameters (Params, K=thousand, M=million), Top-1 and Top-3 Accuracy (Top-1A, Top-3A), Distance-Based Accuracy (DBA), Precision (P), Top-3 Beams (Top-3B), and Top-1 Power Ratio (Top-1PR). Results are averaged over 10 train/test trials.}}
\vspace{-2mm}
\label{ResultsTable}
\resizebox{\columnwidth}{!}{%
\begin{tabular}{|c|c|c|c|c|c|c|}

\hline
\textbf{Model} & \textbf{Params$\downarrow$} & \textbf{Top-[1A,3A]~$\uparrow$} & \textbf{DBA~$\uparrow$} & \textbf{P~$\uparrow$} & \textbf{Top-3B~$\uparrow$} & \textbf{Top-1PR~$\uparrow$} \\ \hline

V
& 389K & 0.47, 0.85 & 0.90 & 0.42 & 0.85 & 0.98 \\ \hline

G
& 25K & 0.45, 0.82 & 0.89 & 0.33 & 0.81 & 0.97 \\ \hline

L
& 389K & 0.45, 0.82 & 0.88 & 0.40 & 0.82 & 0.97 \\ \hline

\shortstack{Early LV\\ \footnotesize no RT}
& 390K & 0.46, 0.83 & 0.89 & 0.40 & 0.83 & 0.98 \\ \hline

\shortstack{Early LV\\ \footnotesize RT\textsubscript{M}}
& 390K & 0.48, 0.85 & 0.90 & 0.42 & 0.84 & 0.98 \\ \hline

\shortstack{Early LVG\textsuperscript{*} \\ \footnotesize no RT}
& 390K & 0.49, 0.87 & 0.91 & 0.43 & 0.87 & 0.98  \\ \hline

\shortstack{Early LVG\textsuperscript{*} \\ \footnotesize RT\textsubscript{M}}
& 390K & 0.47, 0.85 & 0.90 & 0.38 & 0.85 & 0.98  \\ \hline

\shortstack{Late LV\\ \footnotesize AS + RT\textsubscript{M}}
& 807K & 0.48, 0.85 & 0.90 & 0.41 & 0.86 & 0.98  \\ \hline

\shortstack{Late VG\\ \footnotesize no AS/RT}
& 444K & 0.48, 0.86 & 0.90 & 0.39 & 0.85 & 0.98 \\ \hline

\shortstack{Late VG \\ \footnotesize AS + no RT}
& 452K & 0.49, 0.87 & 0.91 & 0.43 & 0.87 & 0.98 \\ \hline

\shortstack{Late VG  \\ \footnotesize AS + RT\textsubscript{M}}
& 452K & 0.49, 0.87 & 0.91 & 0.42 & 0.87 & 0.98 \\ \hline

\shortstack{Late LVG\\ \footnotesize no AS/RT}
& 837K & 0.48, 0.86 & 0.90 & 0.39 & 0.86 & 0.98 \\ \hline

\shortstack{Late LVG\\ \footnotesize AS + no RT}
& 850K & 0.48, 0.87 & 0.91 & 0.43 & 0.88 & 0.98 \\ \hline

\shortstack{Late LVG\\ \footnotesize AS + RT\textsubscript{M}}
& 850K & \textbf{0.50, 0.87} & \textbf{0.91} & \textbf{0.45} & \textbf{0.88} & 0.98 \\ \hline

\shortstack{Late LVG\\ \footnotesize AS + RT\textsubscript{M(1/2)}}
& 850K & 0.49, 0.87 & 0.91 & 0.44 & 0.87 & 0.98 \\ \hline

\shortstack{Late LVG\\ \footnotesize AS + RT\textsubscript{G}}
& 850K & 0.49, 0.86 & 0.91 & 0.45 & 0.88 & 0.98  \\ \hline

\end{tabular}%
}
\vspace{0.5mm}
\begin{minipage}{\columnwidth}
\footnotesize
\textsuperscript{*} Normalization applied before training to the fused dataset for all models with \textsuperscript{*}, excluding G for those containing it.
\end{minipage}
\vspace{-2.5mm}
\end{table}

\begin{figure}[!t]
    \vspace{-3mm}
    \centering
    \includegraphics[width=0.5\textwidth,height=0.45\textheight,keepaspectratio]{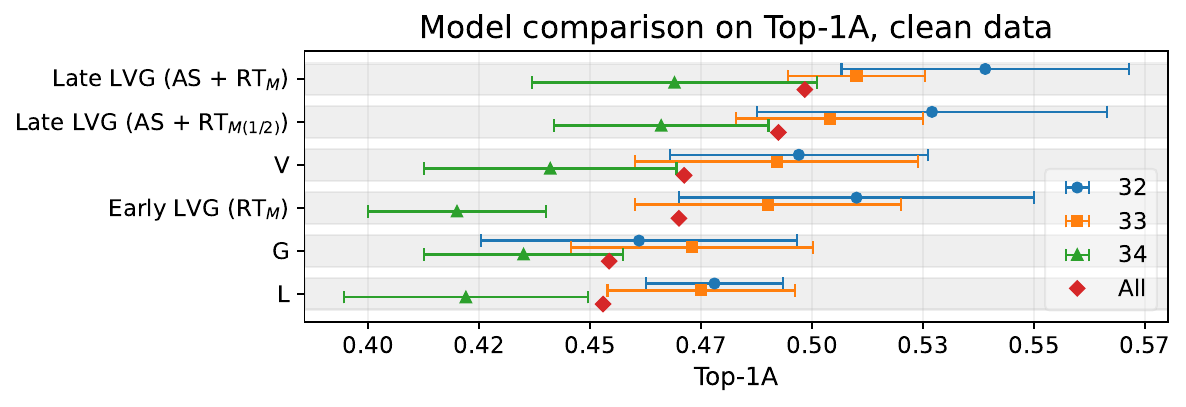}
    \vspace{-7mm}
    \caption{\small{Top-1A comparison of single-modality L, V, and G models and the fused LVG models across individual scenarios (32, 33, 34) and the scenario-weighted average.}
    } 
\vspace{-4mm}
    \label{fig:ModelComparison}
\end{figure}

Among the single-modality models, V achieves the best performance, benefiting from its rich visual context. G and L provide comparable accuracy, although L yields higher precision.

For early fusion, Early LV (no RT) performs slightly worse than the single-modality V model, indicating that early fusion not only fails to exploit the additional L information but can also slightly degrade performance. 

With RT\textsubscript{M}, Early LV improves to a level close to single V, suggesting that RT\textsubscript{M} partially mitigates ineffective fusion learning. However, this model still does not show a clear positive contribution from L, and its performance remains largely limited by V. Early LVG (no RT) improves over the single-modality models and Early LV (no RT). Adding RT\textsubscript{M} to Early LVG slightly lowers clean-condition performance, but its primary benefit is expected under poor-sensor conditions rather than the full-sensor setting, as shown later in Fig.~\ref{fig:EarlyLVG_LateVG_LateLVG}.

For late-fusion models, most configurations perform competitively, with Late LVG (AS+RT\textsubscript{M}) achieving the best overall performance. Late LVG (AS+RT\textsubscript{G}) also preserves high performance in the full-sensor setup. Notably, in late fusion, adding RT does not degrade full-sensor performance, despite randomly removing modality information during training.

From Table~\ref{ResultsTable}, high top-1PR values show that models can select beams with received power close to the optimal beam power under clean test conditions. This agrees with prior observations~\cite{KatSabaMMBP1} that multiple beams may provide similar link quality, highlighting the value of power-based metrics for practical BP evaluation. However, top-KA, DBA, and P still show room for improvement in exact prediction. More importantly, high clean-condition link quality does not ensure reliability under sensor degradation or failure. Therefore, the proposed multimodal method aims not only to improve clean-condition performance but also to maintain robustness by exploiting complementary information under modality degradation.

To further examine scenario-wise behavior, Fig.~\ref{fig:ModelComparison} complements Table~\ref{ResultsTable} by comparing RT\textsubscript{M}-based fusion strategies and the related single-modality models across individual scenarios, with models ordered from top to bottom by overall top-1A. Late LVG (AS+RT\textsubscript{M}) achieves the highest top-1A in all scenarios. 
Scenario~32 is recorded during daytime and provides better RGB visibility, whereas Scenarios~33 and~34 are nighttime scenes, which may reduce visual clutter or yield more uniform illumination. In addition, Scenarios~32 and~33 correspond to the same location under different lighting conditions, allowing the model to benefit from more samples with similar geometry, while Scenario~34 is collected at a different location.
Table~\ref{ResultsTable} and Fig.~\ref{fig:ModelComparison} show that Late LVG consistently outperforms Late VG, Early LVG, and the single-modality baselines under clean-data conditions. Therefore, we adopt Late LVG (AS) as the base architecture for evaluating the impact of the proposed RT\textsubscript{G} configuration.

\vspace{-2mm}
\subsection{Effect of AS}
Table~\ref{ResultsTable} shows that AS improves inference reliability relative to the corresponding models without AS. Before introducing any data degradation, adding AS increases the top-3A and DBA scores of both Late VG and Late LVG by \(1\%\), and improves top-3B by \(2\%\) for both models, indicating that AS strengthens the final fused decision. Although top-1A only evaluates whether the top-ranked beam is correct, P is more sensitive to false-positive beam assignments across classes. Therefore, the \(4\%\) improvement in P for both models, despite only a \(0\)--\(1\%\) gain in top-1A, suggests that AS reduces incorrect class assignments.

\vspace{-2mm}
\subsection{RT\textsubscript{M} Analysis}
\begin{figure}[!t]
    \vspace{-4mm}
    \centering
\includegraphics[width=0.35\textwidth,height=0.4\textheight,keepaspectratio]
    {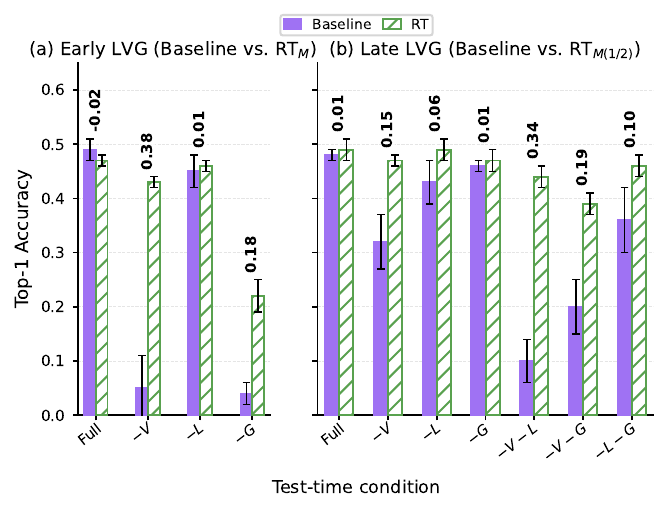}
    \vspace{-3.5mm}
    \caption{\small{Top-1A robustness under full-sensor and missing-modality inference for: (a) Early LVG baseline vs. RT\textsubscript{M}, (b) Late LVG baseline vs. RT\textsubscript{M(1/2)}. Bars: fold means; error bars: standard deviations.}} 
    \label{fig:EarlyLVG_LateVG_LateLVG}
    \vspace{-3mm}
\end{figure}

After establishing the benefit of AS, we evaluate RT\textsubscript{M}. We compare models with and without RT\textsubscript{M}, both trained with AS, under full-sensor and missing-modality test conditions for Early LVG (RT\textsubscript{M}) and Late LVG (AS+RT\textsubscript{M(1/2)}) (Fig.~\ref{fig:EarlyLVG_LateVG_LateLVG}).
The full-sensor setting serves as an empirical upper-bound reference for performance loss under missing modalities. Dropping the V, L, or G modalities is denoted by \(-V\), \(-L\), and \(-G\), respectively. These modality ablations quantify the contribution of each sensor input across different models: a larger drop after removing a modality indicates stronger reliance on that sensor to support the true-beam prediction. 

According to Fig.~\ref{fig:EarlyLVG_LateVG_LateLVG}(a), removing V and G from the Early LVG baseline causes large drops in top-1A, while the RT\textsubscript{M} variant trained with sensor dropout recovers a substantial portion of this loss in each case. Removing L slightly reduces performance for both the baseline and the RT\textsubscript{M} model. 

Fig.~\ref{fig:EarlyLVG_LateVG_LateLVG}(b) provides evidence for the benefit of RT\textsubscript{M(1/2)}. The naive Late LVG baseline, trained without modality dropout, performs poorly when one or two sensors are missing at test time, with the largest degradation occurring when both V and L are unavailable. Its smallest drop occurs when G is removed, indicating lower reliance on G in this setting. In contrast, the RT\textsubscript{M(1/2)} model, remains more stable and achieves higher performance across missing-sensor configurations. Notably, RT\textsubscript{M(1/2)} improves not only robustness under missing modalities but also the full-modality setting by $1\%$.

\begin{figure}[!t]
    \centering
\includegraphics[width=0.5\textwidth,height=0.4\textheight,keepaspectratio]
    {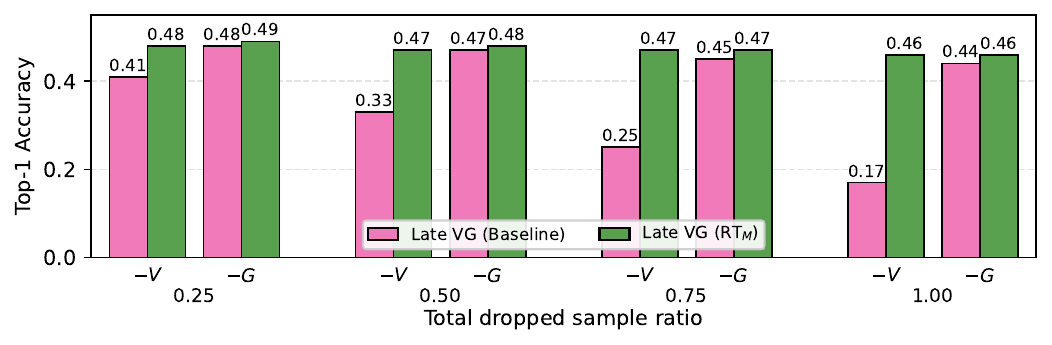}
    \vspace{-6mm}
\caption{\small{Top-1A of Late VG with and without RT\textsubscript{M} under test-time modality dropout applied to different fractions of test samples, with one modality removed at a time.}}
\vspace{-6mm}
\label{fig:alphapercentdrop}
\end{figure}

We further examine the effect of the dropped-sample ratio at test time on Late VG (AS+RT\textsubscript{M}). Rather than degrading the entire test set, modality dropout is applied to different fractions of test samples. As shown in Fig.~\ref{fig:alphapercentdrop}, the baseline model without RT\textsubscript{M} becomes increasingly sensitive as the fraction of samples with missing V grows, while showing lower sensitivity to missing G over the same range. In contrast, the RT\textsubscript{M} variant remains stable across all dropped-sample ratios.

\vspace{-2mm}
\subsection{RT\textsubscript{G} Analysis}
\begin{figure}[!t]
    \vspace{-3.5mm}
    \centering
    \includegraphics[width=0.5\textwidth]{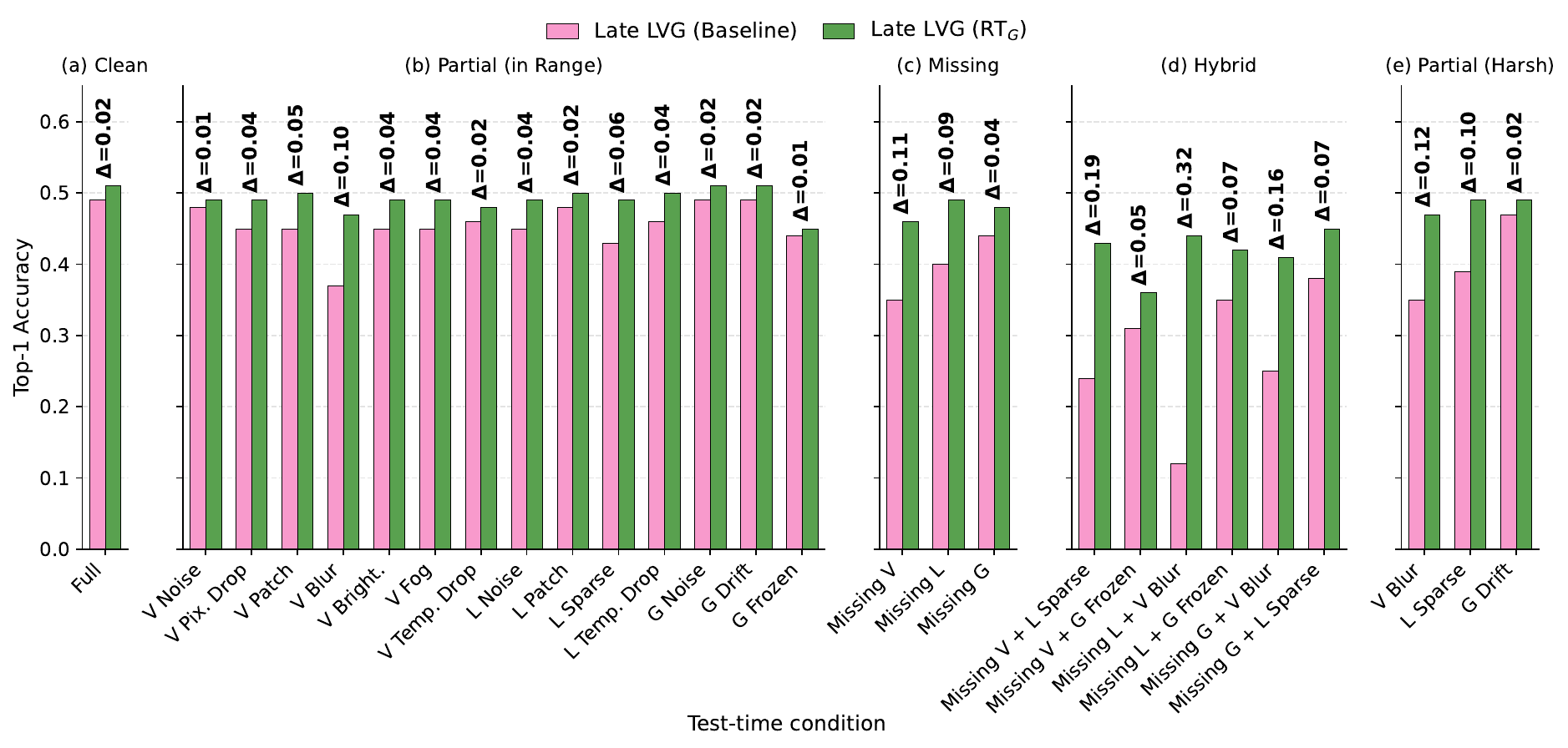}
    \vspace{-7mm}
    \caption{\small{Robustness comparison under clean, missing, partial (in-range intensity), hybrid, and partial (out-of-range intensity) conditions, measured by top-1A.}
    } 
    \label{fig:LateLVG_CleanEntirePartial}
    \vspace{-4mm}
\end{figure}

We further study RT\textsubscript{G} as an alternative to RT\textsubscript{M}, building on Late LVG (AS+RT\textsubscript{M}), which achieved the best performance in the previous simulations. To assess temporal continuity, RT\textsubscript{G} is evaluated over a continuous time interval in which consecutive samples form a realistic sequential driving scenario.
We construct a single train--test split whose test set contains a fixed block of 100 consecutive samples from Scenario~32. A 10-to-90 train--test split was used, consistent with the other models.
Both the no-degradation-training baseline and the RT\textsubscript{G} model are trained on this split and evaluated under clean, missing-modality, in-range partial-degradation, hybrid sensor-failure, and out-of-range harsher partial-degradation conditions. The clean-condition results are reported in Table~\ref{ResultsTable}, while Fig.~\ref{fig:LateLVG_CleanEntirePartial} provides a detailed top-1A breakdown of the RT\textsubscript{G} model and its baseline across the clean setting and each degradation type. Fig.~\ref{fig:LateLVG_CleanEntirePartial}(d) hybrid evaluation is a selected worst-case subset.

Under the clean condition, Fig.~\ref{fig:LateLVG_CleanEntirePartial} shows that RT\textsubscript{G} provides a \(2\%\) improvement over the baseline. Under poor sensor conditions, RT\textsubscript{G} improves robustness across all degradation types relative to the baseline trained without degradation. The largest gain is observed for the Missing L + Blur V hybrid condition, where top-1A improves by 0.32. For the baseline, the most challenging cases are missing V in the missing-modality category, blurred V in the partial vision-degradation category, sparse L in the partial LiDAR-degradation category, frozen G in the partial GPS-degradation category, and missing L + blurred V in the hybrid category. RT\textsubscript{G} also outperforms the baseline for the harsher degradation cases shown in Fig.~\ref{fig:LateLVG_CleanEntirePartial}(e), which means the RT\textsubscript{G} model is robust against severe conditions it has not seen during training too. 

Under the clean condition, Fig.~\ref{fig:LateLVG_CleanEntirePartial} shows that RT\textsubscript{G} improves top-1A by \(2\%\) over the baseline. Under poor-sensor conditions, RT\textsubscript{G} consistently improves robustness across all degradation types relative to the baseline trained without degradation. The largest gain occurs in the Missing L + Blur V hybrid condition, where top-1A increases by 0.32. For the baseline, the most challenging cases are missing V among missing-modality conditions, blurred V among partial vision degradations, sparse L among partial LiDAR degradations, frozen G among partial GPS degradations, and Missing L + Blur V among hybrid conditions. RT\textsubscript{G} also outperforms the baseline under the harsher degradation cases in Fig.~\ref{fig:LateLVG_CleanEntirePartial}(e), indicating robustness to severe conditions not observed during training. These results suggest that training on a sufficiently diverse degradation distribution improves robustness not only within the trained degradation range, but also under harsher or more frequent test-time impairments. This indicates that the model becomes less sensitive to the exact degradation configuration, reducing the need for precise prior knowledge of test-time degradation levels as long as training covers a representative range of sensor impairments.

\vspace{-2mm}
\subsection{Correct-Beam Probability Support}
\begin{figure}[!t]
    \vspace{-1mm}
    \centering
\includegraphics[width=0.5\textwidth,height=0.4\textheight,keepaspectratio]
    {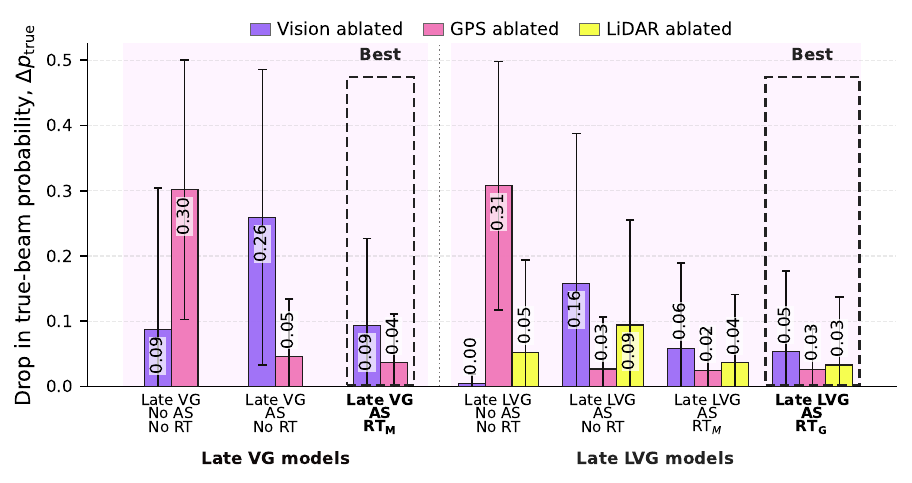}
    \vspace{-7mm}
    \caption{\small{Correct-beam probability support under modality ablation for Late VG/LVG models. Bars report the mean true-beam probability drop over test samples from a selected fold, while error bars denote the corresponding within-sample standard deviation; for RT\textsubscript{G}, the single available fold is used.}} 
    \label{fig:DeltaPTrue}
    \vspace{-2mm}
\end{figure}

\vspace{-2mm}
As a complementary analysis, we examine how adding AS and then either RT\textsubscript{M} or RT\textsubscript{G} improves learning relative to the baseline without AS, RT\textsubscript{M}, or RT\textsubscript{G}. Since late-fusion models outperform the early-fusion variants, we focus on late fusion and measure the decrease in the predicted probability of the ground-truth beam after ablating each modality, as defined in Eq.~\ref{DeltaPTrue} and reported in Fig.~\ref{fig:DeltaPTrue}. This metric provides a finer-grained view of how strongly each modality supports the correct beam, independent of whether the final predicted beam changes. For each model, each bar shows the average decrease in true-beam probability when one modality is removed; larger values indicate stronger reliance.

Comparing Late VG and Late LVG with and without AS shows that AS changes the modality-contribution pattern for correct-beam probability. In both models, AS reduces over-reliance on G and shifts fusion toward stronger use of V. Among the AS-based models, V provides the strongest support for the correct beam, indicating that it carries meaningful beam-selection information, while L and G contribute mainly as complementary modalities. 

The degradation-trained models show smaller and more uniformly distributed drops in true-beam probability, indicating reduced dependence on any single modality and improved robustness to modality removal. Among them, Late LVG (AS+RT\textsubscript{G}) and Late LVG (AS+RT\textsubscript{M}) achieve the smallest and most balanced ablation drops, suggesting the lowest sensitivity to missing modalities. In contrast, the larger standard deviations of models without AS and RT\textsubscript{M} indicate less consistent modality reliance, where ablating the same modality can affect samples differently. Adding AS and RT\textsubscript{M} reduces this variability across ablation cases, leading to more uniform modality contributions and a more balanced fusion strategy. 

\vspace{-3mm}
\subsection{Continuous Temporal Link Reliability}
To evaluate the temporal link reliability of the best model with RT\textsubscript{G}, we use 100 consecutive clean test samples. We employ a sample-level top-3 correctness indicator: a sample is correct if its ground-truth beam is among the model's top-three predictions, and incorrect otherwise. This yields a binary correctness sequence over time, where consecutive correct predictions indicate uninterrupted beam-connectivity intervals and incorrect predictions correspond to link outages. We report the top-3A over this 100-sample block, the longest continuous run of correct predictions, and the number of outages. The no-degradation baseline achieves \(69\%\) top-3A, with a longest continuous correct run of 16 samples. In contrast, the RT\textsubscript{G} model achieves \(76\%\) top-3A, extends the longest continuous correct run to 32 samples. These results show that RT\textsubscript{G} improves the temporal stability and continuity of beam connectivity.

\begin{table}[t]
\centering
\caption{\small{Temporal beam-connectivity reliability of Late LVG (AS+RT\textsubscript{G}) over a consecutive 100-sample test block.}
}
\vspace{-1.5mm}
\label{tab:temporal_link_reliability}
\scriptsize
\setlength{\tabcolsep}{4pt}
\renewcommand{\arraystretch}{0.95}
\resizebox{0.75\columnwidth}{!}{%
\begin{tabular}{|l|c|c|}
\hline
\textbf{Model} & \textbf{Top-3A}$\uparrow$ & \textbf{Longest Successful Run}$\uparrow$  \\
\hline
Late LVG (AS) & 69\% & 16  \\
\hline
Late LVG (AS+RT\textsubscript{G}) & 76\% & 32  \\
\hline
\end{tabular}%
}
\vspace{-4mm}
\end{table}

\vspace{-2mm}
\subsection{Benchmark Comparison}
We benchmark our models against three state-of-the-art multimodal BP methods, as summarized in Table~\ref{tab:Benchmark}. First, we consider the VG Transformer of Tian et al.~\cite{tian2023multimodal}, which was reported to outperform the 2022 challenge leaderboard solution, \textit{BeamBench}~\cite{BeamBench}. We reimplemented this Transformer using the authors' released hyperparameters~\cite{github_t} and evaluated it on our clean dataset using the same 10-fold split. Despite its substantially higher complexity, it underperforms all of our models, indicating a less favorable accuracy--efficiency trade-off. \textit{BeamBench}~\cite{BeamBench}, the winner of the 2022 Multi-Modal Beam Prediction Challenge\footnote{\url{https://www.deepsense6g.net/ml-task-multi-modal-beam-prediction/}.}, uses convolutional autoencoders for multimodal feature fusion. Since its reported results include different scenario coverage, we compute its DBA over Scenarios~32--34 using the reported sample counts of 54, 140, and 131, while excluding Scenario~31. Under this comparison, our best overall model improves DBA by an absolute margin of 0.14. We also compare against \textit{ViT-S/16-Fusion YOLOv8}~\cite{10577431}, which fuses camera, LiDAR, and GPS information using YOLOv8-based segmentation and a vision-transformer encoder. Using the reported DBA values over Scenarios~32--34 with equal weighting, our best model slightly exceeds its recomputed DBA while requiring far fewer parameters. 
Yao et al.'s attention-based missing-modality BP method~\cite{11089951}, evaluated on DeepSense Scenario~9, reports lower top-5A than our results. While their setup considers missing-modality robustness with limited noise injection, their main contribution focuses on missing modalities. In contrast, our study jointly addresses missing modalities and structured sensor degradations. 
Although Yao et al. do not report the total parameter count or YOLOv10 scale, we estimated a lower bound based on the lightest YOLOv10-N model, excluding the remaining modules.
Since AMBER~\cite{wen2026} does not report the exact held-out sample counts after the train--test split, we compute its top-3A over Scenarios~32--34 as the unweighted mean of the reported per-scenario all-input results. While AMBER primarily targets missing modalities, our evaluation covers a broader set of practical sensing failures.
Compared with AMBER, our framework uses fewer input sources and a simpler non-Transformer fusion design. AMBER uses historical beam indices as an additional input stream, while we use beam indices only as supervision labels. 
They do not report total parameter count, but we estimate a lower bound based on their use of ResNet34 image encoder and two ResNet18-based LiDAR encoders.

\begin{table}[t]
\centering
\caption{\small{Benchmark comparison under clean conditions.}}
\vspace{-1.5mm}
\label{tab:Benchmark}
\footnotesize
\setlength{\tabcolsep}{4pt}
\renewcommand{\arraystretch}{0.95}
\resizebox{0.95\columnwidth}{!}{%
\begin{tabular}{|l|c|c|c|c|c|c|}
\hline
\textbf{Model} & \textbf{Miss. Mod.} & \textbf{Partial Degrad.} &\textbf{DBA}$\uparrow$ & \textbf{Top-3A}$\uparrow$ & \textbf{Top-5A}$\uparrow$ & \textbf{Params}$\downarrow$  \\
\hline
Transformer (VG) \cite{tian2023multimodal} & $\times$ & $\times$ & 0.87 & -- & -- & 55.5M  \\
\hline
BeamBench (VG) \cite{BeamBench} & $\times$ & $\times$ & 0.77 & -- & -- & -- \\
\hline
ViT-S/16-Fusion YOLOv8 (LVG) \cite{10577431} & $\times$ & $\times$ & 0.89 & --& -- & 28.4M \\
\hline
Attention-based (LVGR) \cite{11089951} & $\checkmark$ & $\times$ & --& -- & 0.85 & $>$2M \\
\hline
AMBER (LVGR+BeamIdx) \cite{wen2026} & $\checkmark$ & $\times$ & -- & \textbf{0.87} & -- & $>$43M \\
\hline
Ours: Late LVG (AS+RT\textsubscript{M}) & $\checkmark$ & $\times$ & \textbf{0.91} & \textbf{0.87} & \textbf{0.97} & \textbf{850K}  \\
Ours: Late LVG (AS+RT\textsubscript{G}) & $\checkmark$ & $\checkmark$ & \textbf{0.91} & 0.86 & \textbf{0.97} & \textbf{850K}  \\
\hline
\end{tabular}%
}
\vspace{-4mm}
\end{table}

\vspace{-2mm}
\subsection{Efficiency Analysis} 
We evaluate the computational efficiency of the proposed BP models in terms of parameter complexity and inference latency, both critical for real-time mmWave systems. Our choice of CNN-based architectures is motivated by their strong inductive biases, spatial locality, translation invariance, and hierarchical feature extraction, which enable robust generalization even with moderate data volumes \cite{bai2021transformersrobustcnns}. In contrast, transformers rely on self-attention to capture long-range dependencies with minimal prior assumptions, demanding larger datasets and incurring quadratic time and memory costs relative to input size \cite{bai2021transformersrobustcnns}. Given our dataset’s scale and structure, CNN-based models offer superior accuracy–efficiency trade-offs, making them well-suited for BP. 

\subsubsection{Model Complexity}
Model complexity is evaluated by trainable parameters, memory requirements, and channel handling. Trainable parameters, i.e., the weights and biases updated during training, determine model capacity while also affecting computational load, memory usage, and inference speed. Table~\ref{ResultsTable} reports the parameter counts of all models. Single-modality models are generally the most efficient because they process only one data stream; among them, G is the simplest due to its small input size and absence of CNN layers. Early-fusion models have similar parameter counts, whereas late-fusion architectures are more computationally demanding because each modality is processed by a dedicated branch. 

\subsubsection{Inference time}
Real-time BP requires inference within the beam coherence time to ensure validity. Using a conservative channel coherence time of \(\sim143\)~ms from \cite{7742901}, our best Late LVG (AS+RT\textsubscript{G}) model achieves an inference time of 25~ms on CPU\footnote{Experiments were conducted on a dual-socket Intel Xeon Gold 6346 server (3.10\,GHz, 32 cores / 64 threads) with 1\,TB RAM.}, well below channel coherence time and the longer beam coherence time in vehicular mmWave systems. This demonstrates suitability for real-time deployment, with additional latency reduction achievable via GPU acceleration.

\vspace{-2mm}
\subsection{Discussion} 
\vspace{-1mm}
\subsubsection{Superiority of multimodal DL-based BP}
We use DL to reduce the overhead of conventional codebook-based beam training. Classical non-AI methods typically rely on exhaustive or hierarchical RF beam sweeping. In contrast, the proposed model predicts the most likely beam candidates directly from sensory context. Using exhaustive beam sweeping as an overhead reference, a codebook with $N_B$ beams requires probing all \(N_B\) candidate beams. The proposed DL-based scheme reduces this overhead to K beams in Top-K refinement mode. Accordingly, the normalized beam-training overhead decreases from \(1\) to \(K/N_B\), yielding an overhead reduction of \(1-K/N_B\).

Multimodal sensing improves BP reliability by leveraging environmental awareness from vision, LiDAR, and GPS, instead of relying only on RF probing. The gains of multimodal fusion over single-modality baselines confirm the value of combining complementary sensory information. Although early- and late-fusion models achieve strong clean-dataset performance, the analysis of Early LV, for both baseline and robustness-trained models, reveals undesirable over-reliance on vision and inefficient multimodal utilization. Therefore, Late LVG (AS+RT\textsubscript{G}) is the most robust and best-performing model. Overall, it achieves comparable or superior performance to the proposed variants on clean and degraded datasets, while being substantially smaller than state-of-the-art models, yielding the best accuracy--efficiency trade-off.

\subsubsection{Benefits of the proposed robustness approach} 
Sensor robustness is critical for real-world multimodal deployment, where inputs may be incomplete, noisy, partially degraded, or unavailable due to sensor damage, temporary obstruction, environmental interference, or random failures, and fusion models may otherwise over-rely on dominant modalities. The proposed AS+RT strategies improve BP performance from four complementary perspectives. 
1) They preserve, and in some cases slightly improve, clean-sensor performance. 
2) Under partial or complete modality degradation, they maintain satisfactory BP performance and produce low, balanced mean drops in correct-beam probability, with reduced standard deviations under missing-modality ablation. 
3) The RT\textsubscript{G} model also tolerates the out-of-range harsh condition, indicating that training with a sufficiently diverse degradation distribution improves robustness not only within the trained degradation range, but also under harsher test-time degradations. This suggests that the model becomes less sensitive to the exact degradation configuration when training exposes it to a representative range of sensor impairments.
4) The fewer outages and increased longest successful run length over a continuous sample sequence demonstrate longer uninterrupted prediction intervals and more reliable temporal operation. 
Overall, these trends show that AS+RT improves multimodal feature utilization, reduces single-sensor dependence, and stabilizes fused predictions under modality loss. Since these degradation strategies are model-agnostic, they can also be applied to other wireless communication and autonomous-driving settings to reduce sensor coupling and improve robustness.

\subsubsection{Deployment scope}
The proposed framework is intended for outdoor mmWave BP settings with sensing configurations, mobility patterns, and antenna/codebook structures similar to DeepSense6G. Its robustness is distribution-dependent: reliable performance is expected only when the target channel and sensor conditions are sufficiently represented during training. Since BP accuracy can vary with blockage severity, propagation geometry, received-signal quality, and sensor impairments, direct deployment in substantially different settings, including indoor environments, rich-scattering channels, different carrier frequencies, different antenna/codebook designs, and multi-user or spatial-multiplexing systems, requires domain adaptation and additional validation.

\vspace{-2mm}
\section{Conclusion}
\label{Conclusion}
To address real-time sensor failures degrading multimodal mmWave BP, this paper presented a robust framework based on vision, LiDAR, and GPS sensing. The proposed auxiliary supervision and robustness training strategies improve multimodal fusion by promoting discriminative modality representations, better utilization of available sensor information, and reduced dependence on individual modality. Degradation-aware training preserves, and sometimes improves, clean-data performance while maintaining satisfactory BP accuracy under partial and complete sensor degradation. The proposed model keeps the correct-beam prediction probability high even when individual modalities are unavailable, and improves temporal beam-link continuity by sustaining longer successful user--BS connection runs with fewer outages. These gains are achieved in a low-complexity system, especially for late fusion, showing that the model-agnostic RT can reduce sensor coupling and improve robustness in wireless communication and autonomous-driving applications. Future work will explore more advanced robustness mechanisms, including adaptive robustness blocks that can be integrated into existing systems to improve generalizability and applicability.

\section*{Acknowledgment}
The authors acknowledge using ChatGPT for language enhancement, not for generating original scientific content.

\vspace{-2mm}
\bibliographystyle{IEEEtran}
\bibliography{refs}

@ARTICLE{9743549,
  author={Wu, Shunyao and Alrabeiah, Muhammad and Chakrabarti, Chaitali and Alkhateeb, Ahmed},
  journal={IEEE Open Journal of the Communications Society}, 
  title={Blockage Prediction Using Wireless Signatures: Deep Learning Enables Real-World Demonstration}, 
  year={2022},
  volume={3},
  number={},
  pages={776-796},
  doi={10.1109/OJCOMS.2022.3162591}}

@INPROCEEDINGS{9771835,
  author={Charan, Gouranga and Osman, Tawfik and Hredzak, Andrew and Thawdar, Ngwe and Alkhateeb, Ahmed},
  booktitle={2022 IEEE Wireless Communications and Networking Conference (WCNC)}, 
  title={Vision-Position Multi-Modal Beam Prediction Using Real Millimeter Wave Datasets}, 
  year={2022},
  volume={},
  number={},
  pages={2727-2731},
  doi={10.1109/WCNC51071.2022.9771835}}

@inproceedings{va2017position,
  title={Position-aided millimeter wave V2I beam alignment: A learning-to-rank approach},
  author={Va, Vutha and Shimizu, Takayuki and Bansal, Gaurav and Heath, Robert W},
  booktitle={2017 IEEE 28th Annual International Symposium on Personal, Indoor, and Mobile Radio Communications (PIMRC)},
  pages={1--5},
  year={2017},
  organization={IEEE}
}

@inproceedings{wu2022lidar,
  title={{L}i{DAR}-aided mobile blockage prediction in real-world millimeter wave systems},
  author={Wu, Shunyao and Chakrabarti, Chaitali and Alkhateeb, Ahmed},
  booktitle={2022 IEEE Wireless Communications and Networking Conference (WCNC)},
  pages={2631--2636},
  year={2022},
  organization={IEEE}
}

@misc{charan2022multimodal,
      title={Multi-Modal Beam Prediction Challenge 2022: Towards Generalization}, 
      author={Gouranga Charan and Umut Demirhan and João Morais and Arash Behboodi and Hamed Pezeshki and Ahmed Alkhateeb},
      year={2022},
      eprint={2209.07519},
      archivePrefix={arXiv},
      primaryClass={eess.SP}
}

@ARTICLE{alkhateeb2023deepsense,
  author={Alkhateeb, Ahmed and Charan, Gouranga and Osman, Tawfik and Hredzak, Andrew and Morais, Joao and Demirhan, Umut and Srinivas, Nikhil},
  journal={IEEE Communications Magazine}, 
  title={DeepSense 6{G}: A Large-Scale Real-World Multi-Modal Sensing and Communication Dataset}, 
  year={2023},
  volume={61},
  number={9},
  pages={122-128},
}

@ARTICLE{BeamBench,
  author={Arnold, Maximilian and Charan, Gouranga and Demirhan, Umut and Alkhateeb, Ahmed and Alloulah, Mohammed},
  title={Analysis of Multi-Modal Beam Prediction under Distribution Shift}, 
  url = {https://github.com/ITU-AI-ML-in-5G-Challenge/BeamBench},
  year={2022},
  doi={10.1109/TMM.2022.3145663}
}

@article{tian2023multimodal,
  title={Multimodal Transformers for Wireless Communications: A Case Study in Beam Prediction},
  author={Tian, Yu and Zhao, Qiyang and Boukhalfa, Fouzi and Wu, Kebin and Bader, Faouzi and others},
  journal={ITU Journal on Future and Evolving Technologies},
  year={2023},
  volume={4},
  number={3},
  pages={461-471},
}

@article{reviewNathan,
author = {Gaw, Nathan and Yousefi, Safoora and Reisi Gahrooei, Mostafa},
title = {Multimodal data fusion for systems improvement: A review},
journal = {IISE Transactions},
volume = {54},
number = {11},
pages = {1098-1116},
year = {2022},
publisher = {Taylor & Francis},
doi = {10.1080/24725854.2021.1987593},
URL = {https://doi.org/10.1080/24725854.2021.1987593},
eprint = {https://doi.org/10.1080/24725854.2021.1987593}
}

@ARTICLE{9764610,
  author={Salehi, Batool and Reus-Muns, Guillem and Roy, Debashri and Wang, Zifeng and Jian, Tong and Dy, Jennifer and Ioannidis, Stratis and Chowdhury, Kaushik},
  journal={IEEE Transactions on Vehicular Technology}, 
  title={Deep Learning on Multimodal Sensor Data at the Wireless Edge for Vehicular Network}, 
  year={2022},
  volume={71},
  number={7},
  pages={7639-7655},
  doi={10.1109/TVT.2022.3170733}}

@article{alrabeiah2020deep,
  title={Deep learning for mmwave beam and blockage prediction using Sub-6{GH}z channels},
  author={Alrabeiah, Muhammad and Alkhateeb, Ahmed},
  journal={IEEE Transactions on Communications},
  year={2020},
  publisher={IEEE}
}

@inproceedings{alrabeiah2020millimeter,
  title={Millimeter Wave Base Stations with Cameras: Vision-Aided Beam and Blockage Prediction},
  author={Alrabeiah, Muhammad and Hredzak, Andrew and Alkhateeb, Ahmed},
  booktitle={2020 IEEE 91st Vehicular Technology Conference (VTC2020-Spring)},
  pages={1--5},
  year={2020},
  organization={IEEE}
}

@ARTICLE{Alkhateeb1,
  author={W. {Xu} and F. {Gao} and S. {Jin} and A. {Alkhateeb}},
  journal={IEEE Wireless Communications Letters}, 
  title={{3D} Scene-Based Beam Selection for mm{W}ave Communications}, 
  year={2020},
  volume={9},
  number={11},
  pages={1850-1854},
  doi={10.1109/LWC.2020.3005983}}

@inproceedings{sarker2021vision,
  title={Vision-Aided Beam Allocation for Indoor mm{W}ave Communications},
  author={Sarker, Md Abdul Latif and Orikumhi, Igbafe and Kang, Jeongwan and Jwa, Hye-Kyung and Na, Jee-Hyeon and Kim, Sunwoo},
  booktitle={2021 International Conference on Information and Communication Technology Convergence (ICTC)},
  pages={1403--1408},
  year={2021},
  organization={IEEE}
}

@article{nishio2021wireless,
  title={When wireless communications meet computer vision in beyond {5G}},
  author={Nishio, Takayuki and Koda, Yusuke and Park, Jihong and Bennis, Mehdi and Doppler, Klaus},
  journal={IEEE Communications Standards Magazine},
  year={2021},
  volume={5},
  number={2},
  pages={76-83},
  doi={10.1109/MCOMSTD.001.2000047}
}

@article{ahn2022machine,
  title={Machine Learning-based Vision-aided Beam Selection for mm{W}ave Multi-User MISO System},
  author={Ahn, Hyemin and Orikumhi, Igbafe and Kang, Jeongwan and Park, Hyunwoo and Jwa, Hyekyung and Na, Jeehyeon and Kim, Sunwoo},
  journal={IEEE Wireless Communications Letters},
  year={2022},
  publisher={IEEE}
}

@INPROCEEDINGS{morais2022position,
  author={Morais, João and Bchboodi, Arash and Pezeshki, Hamed and Alkhateeb, Ahmed},
  booktitle={ICC 2023 - IEEE International Conference on Communications}, 
  title={Position-Aided Beam Prediction in the Real World: How Useful GPS Locations Actually are?}, 
  year={2023},
  volume={},
  number={},
  pages={1824-1829},
  doi={10.1109/ICC45041.2023.10278998}}

@ARTICLE{Klautau1,
  author={A. {Klautau} and N. {González-Prelcic} and R. W. {Heath}},
  journal={IEEE Wireless Communications Letters}, 
  title={{LIDAR} Data for Deep Learning-Based mm{W}ave Beam-Selection}, 
  year={2019},
  volume={8},
  number={3},
  pages={909-912},}

@ARTICLE{MBTVT,
  author={Zecchin, Matteo and Mashhadi, Mahdi Boloursaz and Jankowski, Mikolaj and Gündüz, Deniz and Kountouris, Marios and Gesbert, David},
  journal={IEEE Transactions on Vehicular Technology}, 
  title={{L}i{DAR} and Position-Aided mm{W}ave Beam Selection with Non-local {CNN}s and Curriculum Training},
  year={2022},
  volume={71},
  number={3},
  pages={2979-2990},
  doi={10.1109/TVT.2022.3142513}}

@ARTICLE{IPC,
  author={Mahdi Boloursaz Mashhadi and Mikolaj Jankowski and Tze-Yang Tung and Szymon Kobus and Deniz Gündüz},
  journal={IEEE Wireless Communications Letters}, 
  title={Federated mm{W}ave Beam Selection Utilizing {LIDAR} Data}, 
  year={2021},
  volume={10},
  number={10},
  pages={2269-2273},
  doi={10.1109/LWC.2021.3099136}}

@INPROCEEDINGS{10464587,
  author={Nie, Jiali and Cui, Yuanhao and Yu, Tiankuo and Mu, Junsheng and Yuan, Weijie and Jing, Xiaojun},
  booktitle={2023 IEEE Globecom Workshops (GC Wkshps)}, 
  title={An Efficient Nocturnal Scenarios Beamforming Based on Multi-Modal Enhanced by Object Detection}, 
  year={2023},
  volume={},
  number={},
  pages={515-520},
  doi={10.1109/GCWkshps58843.2023.10464587}}

@INPROCEEDINGS{10279177,
  author={Gu, Jerry and Salehi, Batool and Pimple, Snehal and Roy, Debashri and Chowdhury, Kaushik R.},
  booktitle={ICC 2023 - IEEE International Conference on Communications}, 
  title={TUNE: Transfer Learning in Unseen Environments for V2X mmWave Beam Selection}, 
  year={2023},
  volume={},
  number={},
  pages={1658-1663},
  doi={10.1109/ICC45041.2023.10279177}}

@INPROCEEDINGS{9796865,
  author={Salehi, Batool and Gu, Jerry and Roy, Debashri and Chowdhury, Kaushik},
  booktitle={IEEE INFOCOM 2022 - IEEE Conference on Computer Communications}, 
  title={FLASH: Federated Learning for Automated Selection of High-band mmWave Sectors}, 
  year={2022},
  volume={},
  number={},
  pages={1719-1728},
  doi={10.1109/INFOCOM48880.2022.9796865}}

@INPROCEEDINGS{9751514,
  author={Reus-Muns, Guillem and Salehi, Batool and Roy, Debashri and Jian, Tong and Wang, Zifeng and Dy, Jennifer and Ioannidis, Stratis and Chowdhury, Kaushik},
  booktitle={2021 17th International Conference on Mobility, Sensing and Networking (MSN)}, 
  title={Deep Learning on Visual and Location Data for V2I mmWave Beamforming}, 
  year={2021},
  volume={},
  number={},
  pages={559-566},
  doi={10.1109/MSN53354.2021.00087}}

@online{github_t,
  author       = {{ITU AI/ML in 5G Challenge}},
  title        = {{DeepSense 6G Dataset}},
  year         = {2024},
  url          = {https://github.com/ITU-AI-ML-in-5G-Challenge/DeepSense6G\_TII},
  note         = {Accessed: 2024-09-22}
}

@INPROCEEDINGS{KatSabaMMBP1,
  author={Vuckovic, Katarina and Hosseini, Saba M. and Rahnavard, Nazanin},
  booktitle={MILCOM 2024 - 2024 IEEE Military Communications Conference (MILCOM)}, 
  title={Revisiting Performance Metrics for Multimodal mmWave Beam Prediction Using Deep Learning}, 
  year={2024},
  volume={},
  number={},
  pages={881-887},
  doi={10.1109/MILCOM61039.2024.10773747}}

@ARTICLE{10912462,
  author={Zhu, Qihao and Wang, Yu and Li, Wenmei and Huang, Hao and Gui, Guan},
  journal={IEEE Transactions on Communications}, 
  title={Advancing Multi-Modal Beam Prediction with Cross-Modal Feature Enhancement and Dynamic Fusion Mechanism}, 
  year={2025},
  volume={},
  number={},
  pages={1-1},
  doi={10.1109/TCOMM.2025.3548021}}

@INPROCEEDINGS{10683225,
  author={Shi, Binpu and Li, Min and Zhao, Ming-Min and Lei, Ming and Li, Liyan},
  booktitle={2024 IEEE 99th Vehicular Technology Conference (VTC2024-Spring)}, 
  title={Multimodal Deep Learning Empowered Millimeter-Wave Beam Prediction}, 
  year={2024},
  volume={},
  number={},
  pages={1-6},
  doi={10.1109/VTC2024-Spring62846.2024.10683225}}

@ARTICLE{10577431,
  author={Tariq, Shehbaz and Arfeto, Brian Estadimas and Khalid, Uman and Kim, Sunghwan and Duong, Trung Q. and Shin, Hyundong},
  journal={IEEE Internet of Things Journal}, 
  title={Deep Quantum-Transformer Networks for Multimodal Beam Prediction in ISAC Systems}, 
  year={2024},
  volume={11},
  number={18},
  pages={29387-29401},
  doi={10.1109/JIOT.2024.3420455}}

@ARTICLE{7742901,
  author={Va, Vutha and Choi, Junil and Heath, Robert W.},
  journal={IEEE Transactions on Vehicular Technology}, 
  title={The Impact of Beamwidth on Temporal Channel Variation in Vehicular Channels and Its Implications}, 
  year={2017},
  volume={66},
  number={6},
  pages={5014-5029},
  doi={10.1109/TVT.2016.2622164}}

@ARTICLE{9791349,
  author={Liu, Chang and Yuan, Weijie and Li, Shuangyang and Liu, Xuemeng and Li, Husheng and Ng, Derrick Wing Kwan and Li, Yonghui},
  journal={IEEE Journal on Selected Areas in Communications}, 
  title={Learning-Based Predictive Beamforming for Integrated Sensing and Communication in Vehicular Networks}, 
  year={2022},
  volume={40},
  number={8},
  pages={2317-2334},
  doi={10.1109/JSAC.2022.3180803}}

@phdthesis{vuckovic2024robust,
  author       = {Vuckovic, Katarina},
  title        = {Robust Machine Learning Based Frameworks for Localization and mmWave Beam Prediction in Massive MIMO Systems},
  school       = {University of Central Florida},
  year         = {2024},
  type         = {Ph.D. Dissertation},
  note         = {Graduate Thesis and Dissertation post-2024, 82},
  url          = {https://stars.library.ucf.edu/etd2024/82}
}

@misc{bai2021transformersrobustcnns,
      title={Are Transformers More Robust Than {CNNs}?}, 
      author={Yutong Bai and Jieru Mei and Alan Yuille and Cihang Xie},
      year={2021},
      eprint={2111.05464},
      archivePrefix={arXiv},
      primaryClass={cs.CV},
      url={https://arxiv.org/abs/2111.05464}, 
}

@misc{wang2025visionlanguagemodelvlmmeets,
      title={When Vision-Language Model {(VLM)} Meets Beam Prediction: A Multimodal Contrastive Learning Framework}, 
      author={Ji Wang and Bin Tang and Jian Xiao and Qimei Cui and Xingwang Li and Tony Q. S. Quek},
      year={2025},
      eprint={2508.00456},
      archivePrefix={arXiv},
      primaryClass={eess.SP},
      url={https://arxiv.org/abs/2508.00456}, 
}

@INPROCEEDINGS{10826962,
  author={Wang, Shiyue and Chen, Wei and Li, Wen and Zhang, Yuanyuan and Sun, Xiaofang and Ai, Bo},
  booktitle={2024 16th International Conference on Wireless Communications and Signal Processing (WCSP)}, 
  title={Spatial-Temporal Beam Prediction: A Deep Learning-Based Approach With Compressed Multi-Beam Measurements}, 
  year={2024},
  volume={},
  number={},
  pages={272-277},
  doi={10.1109/WCSP62071.2024.10826962}}

@InProceedings{pmlr,
  title = 	 {Learning End-to-end Multimodal Sensor Policies for Autonomous Navigation},
  author = 	 {Liu, Guan-Horng and Siravuru, Avinash and Prabhakar, Sai and Veloso, Manuela and Kantor, George},
  booktitle = 	 {Proceedings of the 1st Annual Conference on Robot Learning},
  pages = 	 {249--261},
  year = 	 {2017},
  editor = 	 {Levine, Sergey and Vanhoucke, Vincent and Goldberg, Ken},
  volume = 	 {78},
  series = 	 {Proceedings of Machine Learning Research},
  month = 	 {13--15 Nov},
  publisher =    {PMLR},
}

@inproceedings{mohta2020investigating,
  title     = {Investigating the Effect of Sensor Modalities in Multi-Sensor Detection-Prediction Models},
  author    = {Mohta, Abhishek and Chou, Fang-Chieh and Becker, Brian C. and Vallespi-Gonzalez, Carlos and Djuric, Nemanja},
  booktitle = {Machine Learning for Autonomous Driving Workshop at the 34th Conference on Neural Information Processing Systems (NeurIPS)},
  year      = {2020},
  address   = {Vancouver, Canada},
  note      = {Workshop paper}
}

@article{2670313,
author = {Srivastava, Nitish and Hinton, Geoffrey and Krizhevsky, Alex and Sutskever, Ilya and Salakhutdinov, Ruslan},
title = {Dropout: a simple way to prevent neural networks from overfitting},
year = {2014},
issue_date = {January 2014},
publisher = {JMLR.org},
volume = {15},
number = {1},
issn = {1532-4435},
journal = {J. Mach. Learn. Res.},
month = jan,
pages = {1929–1958},
numpages = {30}
}

@INPROCEEDINGS{10230356,
  author={Holste, Gregory and van der Wal, Douwe and Pinckaers, Hans and Yamashita, Rikiya and Mitani, Akinori and Esteva, Andre},
  booktitle={2023 IEEE 20th International Symposium on Biomedical Imaging (ISBI)}, 
  title={Improved Multimodal Fusion for Small Datasets with Auxiliary Supervision}, 
  year={2023},
  volume={},
  number={},
  pages={1-5},
  doi={10.1109/ISBI53787.2023.10230356}}

@INPROCEEDINGS{9156420,
  author={Wang, Weiyao and Tran, Du and Feiszli, Matt},
  booktitle={2020 IEEE/CVF Conference on Computer Vision and Pattern Recognition (CVPR)}, 
  title={What Makes Training Multi-Modal Classification Networks Hard?}, 
  year={2020},
  volume={},
  number={},
  pages={12692-12702},
  doi={10.1109/CVPR42600.2020.01271}}

@misc{wu2024,
      title={Multimodal Multi-loss Fusion Network for Sentiment Analysis}, 
      author={Zehui Wu and Ziwei Gong and Jaywon Koo and Julia Hirschberg},
      year={2024},
      eprint={2308.00264},
      archivePrefix={arXiv},
      primaryClass={cs.CL},
      url={https://arxiv.org/abs/2308.00264}, 
}

@article{LI2026132193,
title = {A comprehensive review on deep supervision in computer vision},
journal = {Neurocomputing},
volume = {666},
pages = {132193},
year = {2026},
issn = {0925-2312},
doi = {https://doi.org/10.1016/j.neucom.2025.132193},
url = {https://www.sciencedirect.com/science/article/pii/S0925231225028656},
author = {Renjie Li and Xinyi Wang and Guan Huang and Wenli Yang and Kaining Zhang and Xiaotong Gu and Son N. Tran and Saurabh Garg and Jane Alty and Quan Bai},
}

@misc{liao2025benchmarking,
      title={Benchmarking Multi-modal Semantic Segmentation under Sensor Failures: Missing and Noisy Modality Robustness}, 
      author={Chenfei Liao and Kaiyu Lei and Xu Zheng and Junha Moon and Zhixiong Wang and Yixuan Wang and Danda Pani Paudel and Luc Van Gool and Xuming Hu},
      year={2025},
      eprint={2503.18445},
      archivePrefix={arXiv},
      primaryClass={cs.CV},
      url={https://arxiv.org/abs/2503.18445}, 
}

@INPROCEEDINGS{10588664,
  author={Beemelmanns, Till and Zhang, Quan and Geller, Christian and Eckstein, Lutz},
  booktitle={2024 IEEE Intelligent Vehicles Symposium (IV)}, 
  title={MultiCorrupt: A Multi-Modal Robustness Dataset and Benchmark of LiDAR-Camera Fusion for 3D Object Detection}, 
  year={2024},
  volume={},
  number={},
  pages={3255-3261},
  doi={10.1109/IV55156.2024.10588664}}

@ARTICLE{11089951,
  author={Yao, Lebin and Ma, Ke and Wang, Pengyu and Sang, Yiliang and Wang, Zhaocheng},
  journal={IEEE Wireless Communications Letters}, 
  title={Robust Multimodal Beam Prediction With Missing Modality}, 
  year={2025},
  volume={14},
  number={10},
  pages={3264-3268},
  doi={10.1109/LWC.2025.3591611}}

@article{Shankar,
  title={Progressive Fusion for Multimodal Integration},
  author={Shiv Shankar and Laure Thompson and Madalina Fiterau},
  journal={ArXiv},
  year={2022},
  volume={abs/2209.00302},
  url={https://api.semanticscholar.org/CorpusID:251979302}
}

@misc{wen2026,
      title={AMBER: An Adaptive Multimodal Mask Transformer for Beam Prediction with Missing Modalities}, 
      author={Chenyiming Wen and Binpu Shi and Min Li and Ming-Min Zhao and Min-Jian Zhao and Jiangzhou Wang},
      year={2026},
      eprint={2512.11331},
      archivePrefix={arXiv},
      primaryClass={cs.IT},
      url={https://arxiv.org/abs/2512.11331}, 
}

@INPROCEEDINGS{10209007,
  author={Yu, Kaicheng and Tao, Tang and Xie, Hongwei and Lin, Zhiwei and Liang, Tingting and Wang, Bing and Chen, Peng and Hao, Dayang and Wang, Yongtao and Liang, Xiaodan},
  booktitle={2023 IEEE/CVF Conference on Computer Vision and Pattern Recognition Workshops (CVPRW)}, 
  title={Benchmarking the Robustness of LiDAR-Camera Fusion for 3D Object Detection}, 
  year={2023},
  volume={},
  number={},
  pages={3188-3198},
  doi={10.1109/CVPRW59228.2023.00321}}

@inproceedings{hosseini2027blockage,
  author    = {S. M. Hosseini and K. Vuckovic and A. Safari and N. Rahnavard},
  title     = {Integrated Multimodal Blockage Detection and Beam Prediction
               for {mmWave} {V2I} Communications},
  booktitle = {Proc. IEEE SoutheastCon},
  year      = {2027},
  note      = {Accepted, to appear}
}
\end{document}